\documentclass[a4paper,11pt]{article}
\pdfoutput=1 
\usepackage{jcappub} 
\usepackage{enumerate}
\usepackage{graphicx}
\usepackage{epsf,amsmath,bbold,amsfonts,stmaryrd}
\usepackage[utf8]{inputenc}
\usepackage[makeroom]{cancel}
\usepackage{mathrsfs}
\usepackage{appendix}
\usepackage{amssymb}
\usepackage{float}
\usepackage{color} 
\usepackage[normalem]{ulem} 
\usepackage{indentfirst}
\usepackage{url}
\usepackage{float}
\usepackage{mathtools}
\usepackage{verbatim}
\usepackage[colorinlistoftodos]{todonotes}
\usepackage{MnSymbol,wasysym}

\usepackage{ulem} 

\newcommand\Tstrut{\rule{0pt}{2.6ex}}         
\newcommand\Bstrut{\rule[-2.6ex]{0pt}{5pt}}   

\usepackage{footnote}
\makesavenoteenv{tabular}
\makesavenoteenv{table}

\usepackage{booktabs}
\usepackage{siunitx}

\newcommand{\be}{\begin{equation}}
\newcommand{\ee}{\end{equation}}
\def\bea{\begin{eqnarray}}
\def\eea{\end{eqnarray}}

\def\ba{\begin{array}}
\def\ea{\end{array}}

\def\bc{\begin{center}}
\def\ec{\end{center}}

\def\bl{\begin{flushleft}}
\def\el{\end{flushleft}}

\def\br{\begin{flushright}}
\def\er{\end{flushright}}

\def\bi{\begin{itemize}}
\def\ei{\end{itemize}}

\def\bt{\begin{tabular}}
\def\et{\end{tabular}}

\newtheorem{question}{Question}
\def\bq{\begin{question}}
\def\eq{\end{question}}

\newtheorem{definition}{Def}
\def\bd{\begin{definition}}
\def\ed{\end{definition}}

\newtheorem{answer}{Answer}
\def\ban{\begin{answer}}
\def\ean{\end{answer}}

\newtheorem{possibleanswer}{Possible answer}
\def\bpa{\begin{possibleanswer}\normalfont}
\def\epa{\end{possibleanswer}}

\newtheorem{theorem}{Theorem}
\def\bth{\begin{theorem}}
\def\eth{\end{theorem}}

\title{
\huge{Gravitational waves from global cosmic strings in quintessential inflation}}
\author[a]{Dario Bettoni,}
\emailAdd{bettoni@thphys.uni-heidelberg.de}  
\author[a]{Guillem Dom{\`e}nech,}
\emailAdd{domenech@thphys.uni-heidelberg.de}  
\author[a,b]{Javier Rubio}
\emailAdd{javier.rubio@helsinki.fi} 
\vspace{5cm}
\affiliation[a]{Institut f\"ur Theoretische Physik, Ruprecht-Karls-Universit\"at Heidelberg, \\
Philosophenweg 16, 69120 Heidelberg, Germany}
\affiliation[b]{Department of Physics and Helsinki Institute of Physics, \\  PL 64, FI-00014 University of Helsinki, Finland} 

\abstract{The combination of non-minimal couplings to gravity with the post-inflationary kinetic-dominated era typically appearing in quintessential inflation scenarios  may lead to the spontaneous symmetry breaking of internal symmetries and its eventual restoration at the onset of radiation domination. On general grounds, the  breaking of these symmetries leads to the generation of short-lived topological defects that tend to produce gravitational waves until the symmetry is restored. We study here the background of gravitational waves generated by a global cosmic string network following the dynamical symmetry breaking and restoration of a $U(1)$ symmetry. The resulting power spectrum depends on the duration of the heating process and it is potentially detectable, providing a test on the existence of non-minimal couplings to gravity and the characteristic energy scale of post-inflationary physics.} 
\arxivnumber{1810.11117 }
\begin{document}
\maketitle

\section{Introduction} 

The direct detection of gravitational waves (GW) from black hole and neutron star mergers 
\cite{Abbott:2016blz,Abbott:2016nmj,TheLIGOScientific:2017qsa}, together with the observation of an  electromagnetic counterpart \cite{Monitor:2017mdv}, have opened a new window for observational cosmology. Interestingly, these individual two-body sources are just one of the many signals that could be detected by present and future experiments.
A stochastic background of GWs could be produced, for instance, during  inflation and (re)heating \cite{Starobinsky:1979ty,Rubakov:1982df,GarciaBellido:2007af} or be associated with exotic post-inflationary physics such as phase transitions or topological defects \cite{Grojean:2006bp,Weir:2017wfa,Vilenkin:2000jqa,Copeland:2009ga,Dufaux:2010cf,Figueroa:2012kw}.
The discovery of any of these backgrounds would provide an extremely valuable piece of information on energy scales much beyond the reach of any particle physics accelerator.

Topological defects are created whenever there is a spontaneous symmetry breaking, usually associated with some critical temperature. A temporary network of these objects is also  expected in the presence of non-minimal couplings to gravity, provided the existence of a kinetic dominated  era with stiff  equation-of-state parameter $w>1/3$  \cite{Bettoni:2018utf}.
In this paper, we argue that a stochastic background of GWs could be produced by a temporary network of cosmic string appearing in quintessential inflationary models with non-oscillatory runaway potentials. The main ingredient of our proposal is a \textit{subdominant} non-minimally coupled $U(1)$ scalar field on top of the inflaton and matter sectors.
The non-minimal coupling to gravity renders heavy the $U(1)$ field during inflation, preventing the generation of dangerous isocurvature perturbations \cite{Bettoni:2018utf}. 

After inflation, where the violation of the slow-roll conditions together with the absence of a minimum for the inflaton field leads to the onset of a kinetic-dominated era, the Ricci scalar turns negative and with it the effective mass of the $U(1)$ field. The symmetry becomes then spontaneously broken and the field develops a new vacuum at large field values, leading to the formation of a cosmic string network by the standard Kibble mechanism \cite{Copeland:2009ga}.
The oscillations of the strings within the Hubble volume generate a GWs spectrum with an energy density that becomes cumulatively enhanced with respect to the rapidly decreasing energy density of the kinetic-dominated background. The production of GWs continues till the onset of the hot big bang era, where the Ricci scalar vanishes and the negative mass term enforcing the symmetry breaking effectively disappears. When that happens, the initial symmetry is restored and the origin becomes again the absolute minimum of the potential. The complex $U(1)$ field starts spiraling around it, destroying the cosmic string network and halting the GW production. Since the spectrum of GWs is frozen during the long-lasting radiation-dominated era, the final energy density in GWs remains essentially unchanged up to the present cosmological epoch.

This paper is organized as follows. In Section \ref{sec:Q-inflation} we recapitulate the most important aspects of quintessential inflation. The dynamics of a subdominant non-minimally coupled $U(1)$ field in this cosmological background is described in Section \ref{sec:spectator}. Section 
\ref{sec:GW_production} discuss the formation of cosmic strings and the salient features of the associated GW spectrum. Section \ref{sec:GW_today} is devoted to the computation of the present day GW spectrum and to the discussion of the constraints that present and future experiments could cast on the model parameters and early universe physics. Finally, our conclusions are drawn in Section \ref{sec:conclusions}.

\section{Quintessential inflation}\label{sec:Q-inflation}

Inflation and dark energy are usually understood as two independent epochs  in the history of the Universe. Note, however, that there is no fundamental reason for this to be the case. The common features of these cosmological periods could be related, for instance, to some underlying principle, such as the explicit \cite{GarciaBellido:2011de,Karananas:2016kyt,Casas:2017wjh} or emergent realization of scale invariance in the vicinity of non-trivial fixed points \cite{Wetterich:1987fm,Wetterich:1994bg, Wetterich:2014gaa,Rubio:2017gty}. A framework that fits well with the last possibility is quintessential inflation \cite{Peebles:1998qn,Spokoiny:1993kt,Brax:2005uf,Hossain:2014xha,Agarwal:2017wxo,Geng:2017mic,Dimopoulos:2017zvq,Rubio:2017gty}. In its simplest terms, this paradigm makes use of a single degree of freedom --- dubbed \textit{cosmon} \cite{Peccei:1987mm} --- to provide a unified description of inflation and dark energy. Although certain parametrizations involving a non-canonical cosmon field are certainly preferred to highlight the aforementioned connection to scale symmetry \cite{Wetterich:1987fm,Wetterich:1994bg, Wetterich:2014gaa,Rubio:2017gty}, we will follow here the standard approach and describe quintessential inflation in terms of a canonically normalized cosmon field $\phi$ with Lagrangian density 
\begin{equation}\label{eq:cosmon}
\frac{\cal L}{\sqrt{-g}}=\frac{M_{\rm P}^2}{2}R -\frac{1}{2}\partial_\mu \phi \partial^\mu \phi -U(\phi)\,,
\end{equation}
where $M_{\rm P}=(8\pi G)^{-1/2}$ is the reduced Planck mass and $R$ is the Ricci scalar.
A particular model within the paradigm is specified by a choice of the potential $U(\phi)$, which is required to be of the \textit{runaway form} in order to support inflation and dark energy. Several quintessential potentials have been proposed in the literature \cite{Wetterich:1987fm,Wetterich:1994bg,Wetterich:2013jsa,Wetterich:2014gaa,Hossain:2014xha,Rubio:2017gty,Wang:2018kly}, each of them leading to slightly different predictions for the inflationary and dark energy observables. Despite numerical dissimilarities, the background evolution of the Universe is rather insensitive to the precise form of $U(\phi)$ and involves always the same sequence of cosmological epochs. 
At early times, the scalar field is generically displaced at large field values, allowing for inflation with the standard chaotic initial conditions. The inflationary epoch will end, as usual, when the kinetic energy density of the cosmon starts to dominate over the potential counterpart, signaling the break-down of the slow-roll approximation.  In the absence of a potential minimum, the Universe will inevitably enter a \textit{kination} or \textit{deflation} era with effective equation-of-state parameter $w=1$ \cite{Spokoiny:1993kt}. The duration of this unusual cosmological epoch depends on the efficiency of the heating process.

A plethora of heating mechanisms within quintessential inflation have been proposed in the literature \cite{Ford:1986sy,Damour:1995pd,Peebles:1998qn,Felder:1999pv,Feng:2002nb,BuenoSanchez:2007jxm,Rubio:2017gty,Dimopoulos:2018wfg,Nakama:2018gll}. All these mechanisms share a common feature: the decay of the inflaton field does not need to be complete. Even if the initial particle creation is small, the rapid decrease of the cosmon energy density during kinetic domination ($\rho_\phi\sim a^{-6}$) as compared to the redshift of the created plasma $(\rho_{\rm R}\sim a^{-4})$ will eventually lead to radiation domination. A useful way of parametrizing the duration of this transition period is to make use of a \textit{heating efficiency} parameter \cite{Rubio:2017gty}
\begin{equation}\label{thetadef}
\Theta\equiv \frac{\rho_{\rm R}^{\rm kin}}{\rho^{\rm kin}_\phi}=\left(\frac{a_{\rm kin}}{a_{\rm rad}}\right)^2\,,
\end{equation}
where $\rho^{\rm kin}_\phi$ and $\rho_{\rm R}^{\rm kin}$ stand for the energy density of the cosmon and the heating products at the beginning of kinetic domination and $a_{\rm kin}$ and $a_{\rm rad}$ denote respectively the values of the scale factor at the onset of kinetic and radiation domination. For a fiducial Hubble rate $H_{\rm kin }\sim 10^{11}\,{\rm GeV}$,
the typical heating efficiencies per degree of freedom vary between $\Theta\sim 10^{-19}$ in gravitational heating scenarios \cite{Ford:1986sy,Damour:1995pd,Peebles:1998qn} and $\Theta\sim {\cal O}(1)$ in heating scenarios involving matter fields \cite{Felder:1999pv,Rubio:2017gty}.
The minimal value of $\Theta$ is restricted by big bang nucleosynthesis, where the standard hot big bang history should be definitely recovered. In particular, quintessential inflation generates a \textit{primordial} gravitational wave background that becomes blue tilted during kinetic domination \cite{Sahni:1990tx,Rubio:2017gty}, 
\begin{equation}
\Omega_{\rm GW}(k)\sim k \,,
\end{equation}
and may excessively contribute to the effective number of relativistic degrees of freedom at big bang nucleosynthesis, modifying with it the light elements' abundance. The (integrated) nucleosynthesis constraint on the GW density fraction \cite{Maggiore:1999vm,Caprini:2018mtu},
\begin{equation}\label{GWbound1}
h^2\int_{k_{\rm BBN}}^{k_{\rm end}}\Omega_{\rm GW}(k)\, d\ln k \lesssim 1.12\times 10^{-6}\,, 
\end{equation}
with $h=0.678$ and $k_{\rm end}$ and $k_{\rm BBN}$ the momenta associated to the horizon scale at the end  of inflation and at big bang nucleosynthesis, 
translates into a lower bound on the heating efficiency \cite{Rubio:2017gty}
\begin{eqnarray}\label{GWbound2}
&&\Theta\gtrsim          10^{-16}\left(\frac{H_{\rm kin}}{10^{11}\,{\rm GeV}}\right)^2\,,
\end{eqnarray}
which should be compared with the efficiency of a given heating scenario.

Beyond the onset of radiation domination, the background follows the usual hot big bang evolution ~\cite{Rubio:2017gty}. 
At the early stages of radiation domination, the cosmon field  \textit{freezes} to a constant value, allowing for the resurgence of its potential 
energy density. When this component re-approaches the decreasing energy density of the radiation fluid, the evolution of the system settles down to a scaling solution in which the scalar energy density \textit{tracks} the background evolution \cite{Wetterich:2007kr,Amendola:2018ltt}. Eventually the cosmon will exit the tracking regime, leading to a dark energy dominated era ~\cite{Wetterich:2007kr,Amendola:2007yx}.  

\section{Non-minimally coupled spectator field dynamics}\label{sec:spectator}

Let us study the dynamics of a \textit{subdominant} non-minimally coupled $U(1)$ scalar field in the quintessential inflation scenario discussed in the previous section. To this end, we  consider a Lagrangian density
\begin{equation}\label{LBL}
\frac{\mathcal{L}_\chi}{\sqrt{-g}} = -\partial_\mu\chi^\dagger \partial^\mu\chi - \left(m_\chi^2+\xi R\right)\chi^\dagger \chi - \lambda \frac{(\chi^\dagger \chi)^{\frac{n}{2}}}{\Lambda^{n-4}}\,,
\end{equation}
\begin{figure}
\begin{center}
\includegraphics[scale=0.7]{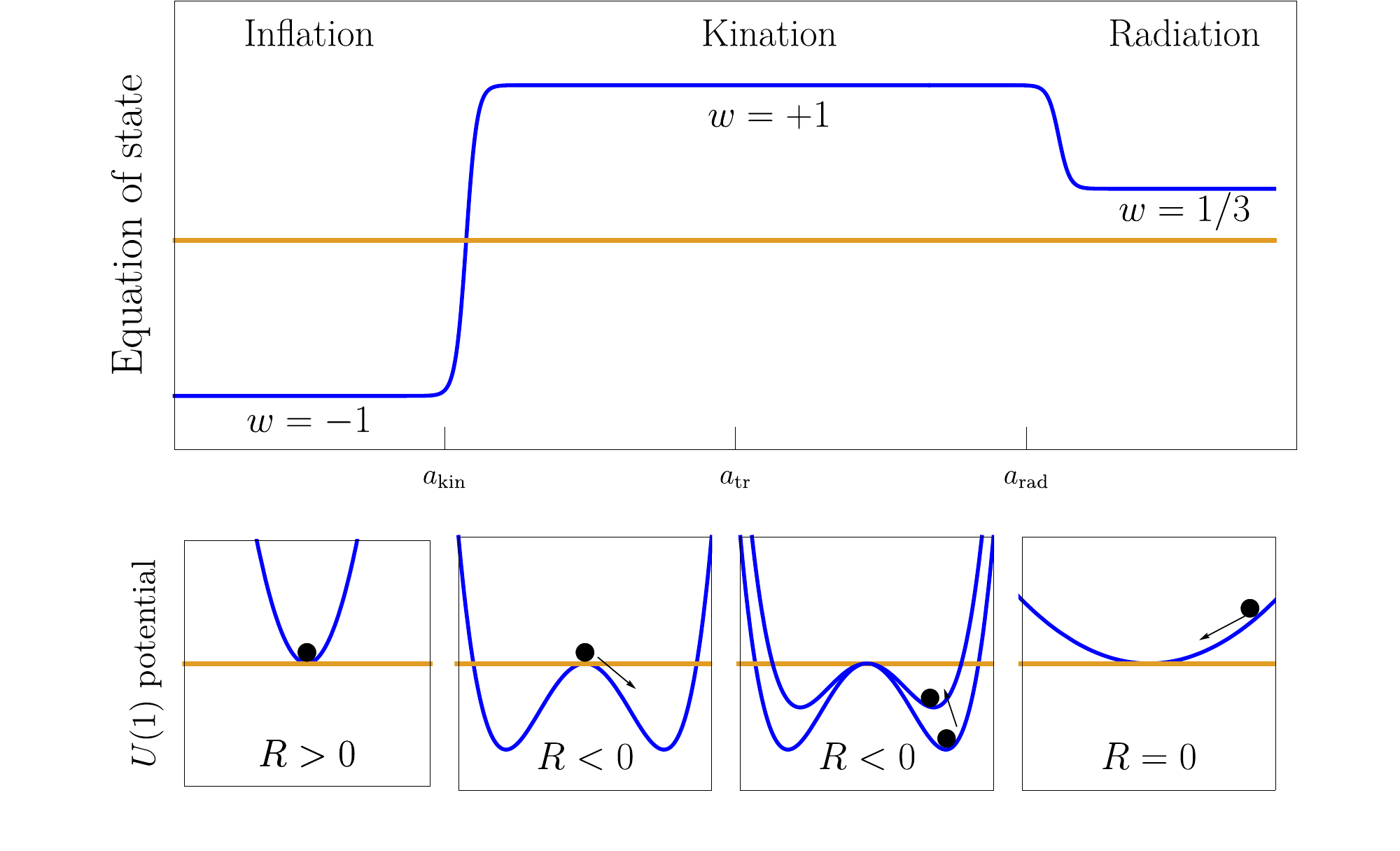}
\caption{Schematic behaviour of the $U(1)$ potential and of the $\chi$ field evolution
during the different cosmological epochs in quintessential inflationary scenarios. The horizontal lines correspond to the zero value of the corresponding quantity.}
\label{fig:sketch}
\end{center}
\end{figure}
with $m_\chi^2$ a bare mass parameter, $\xi$ is a positive dimensionless parameter, $\lambda$ a positive dimensionless coupling, $n\geq 4$ and  $\Lambda$ a cutoff scale (\textit{a priori} unknown). The non-minimal coupling to gravity is motivated by quantum field computations in curved spacetime \cite{Birrell:1982ix} and its associated phenomenology has been widely studied in the literature, with applications in baryogenesis \cite{Bettoni:2018utf}, (re)heating \cite{Dimopoulos:2018wfg} and dark matter production \cite{Fairbairn:2018bsw,Alonso-Alvarez:2018tus}. 

Rather than specifying a particular quintessential inflation potential in Eq.~\eqref{eq:cosmon}, we will follow here a model-independent approach and describe the background evolution of the Universe in  terms  of  the Ricci scalar behaviour at the different cosmological epochs. For a flat Friedmann-Lemaitre-Robertson-Walker \textit{background} metric $g_{\mu\nu}={\rm diag}(-1,a^2(t)\,\delta_{ij})$, the Ricci scalar in Eq.~\eqref{LBL} can be written as 
\begin{equation}
R=3(1-3w) H^2\,,
 \end{equation}
with $H=\dot a/a$ the Hubble rate and $w$ the equation-of-state of  the dominant energy component. 

The evolution of the $U(1)$ potential is summarized in Fig. \ref{fig:sketch}. During the inflationary stage, the parameter $w$ approaches the de Sitter value $w\simeq -1$ and the Ricci scalar is constant and positive,
\begin{equation}\label{Rinf}
R=12 H^2\,.
\end{equation}
The mass term for the $\chi$ field in Eq.~\eqref{LBL} is then also positive definite and the ground state of the system is located at the origin of the effective potential, $ \vert\chi_{\rm min}\vert=0$. This implies, not only the $U(1)$ symmetry is preserved, but also that the $U(1)$ field is heavy during inflation and no sizable isocurvature perturbations are generated \cite{Bettoni:2018utf}.

As soon as the slow-roll conditions are violated, the system enters a deflation period with stiff equation-of-state parameter $w=1$. During this era, the Ricci scalar flips sign 
\begin{equation}\label{Rkin}
R=-6 H^2\,.
\end{equation}
If the associated Hubble-induced mass in Eq.~\eqref{LBL} exceeds the bare mass of the $\chi$ field at that time, the $U(1)$ symmetry becomes spontaneously broken.  
When that happens, the field starts rolling down from the origin to the new minimum of the potential, which is time-dependent and located at
\begin{equation}\label{eq:chimint}
\vert \chi_{\rm min}(a)\vert = \vert \chi_{\rm min}(a_{\rm kin})\vert \left(\frac{a}{a_{\rm kin}}\right)^{-\frac{6}{n-2}}\,, \hspace{10mm} \vert \chi_{\rm min}(a_{\rm kin})\vert \simeq       \left[\frac{12\,\xi}{n\lambda }\left(\frac{H_{\rm kin}}{M_{P}}\right)^2\left(\frac{\Lambda}{M_{P}}\right)^{n-4}\right]^{\frac{1}{n-2}}M_{P}\,.
\end{equation}
 During the first stages of kinetic domination, the higher dimensional operators in  Eq.~\eqref{LBL} can be safely neglected. In this case, the evolution of the field in its way to the time-dependent minimum \eqref{eq:chimint} can be well-described by the approximate solution \cite{Dimopoulos:2018wfg,Bettoni:2018utf}
\begin{equation}\label{eq:chi_roll}
\vert \chi (a) \vert \simeq \vert \chi (a_{\rm kin})\vert  \left(\frac{a}{a_{\rm kin}}\right)^{\sqrt{6\xi}} \,,\hspace{15mm} \vert \chi (a_{\rm kin})\vert  \equiv \frac{1+\sqrt{6\xi}}{2\sqrt{2} \pi} H_{\rm kin}\,,
\end{equation} 
where we have neglected the subleading mass term $m_\chi$ and assumed standard vacuum fluctuations as initial conditions.
An estimate of the number of e-folds $\Delta N_{\rm tr}$  needed to reach \eqref{eq:chimint}  can be obtained by equating $\vert\chi(a_{\rm tr})\vert=\vert\chi_{\rm min}(a_{\rm tr})\vert$ with $a_{\rm tr}$ the value of the scale factor at the transition time,
\begin{equation}\label{eq:DN}
 \Delta N_{\rm tr}\equiv \ln \left(\frac{a_{\rm tr}}{a_{\rm kin}}\right)=\frac{n-2}{(n-2)\sqrt{6\xi}+6}\ln\left[\frac{ \vert \chi_{\rm min}(a_{\rm kin})\vert }{\vert \chi (a_{\rm kin})\vert }\right]\,,
\end{equation}

Once in the minimum, the field tracks its evolution until the $U(1)$ symmetry is restored.  Then, the origin becomes again the ground state of the system and the spectator field $\chi$ starts oscillating around it, provided that the bare mass $m_\chi$ exceeds the instantaneous Hubble rate at the time. If this is not the case, the field will stay frozen at its restoration value until the Hubble parameter has decreased enough to allow for the oscillations. 

\section{Cosmic string network and GW production \label{sec:GW_production}}

At the onset of kinetic domination the phases of the $U(1)$ field are randomly distributed, with a correlation length of the order of the Hubble parameter at that time \cite{Zeldovich:1974uw}. As soon as the horizon expands, multiple patches with different phases enter in causal contact, leading to the formation of a cosmic string network \cite{Kibble:1976sj}. The spontaneous breaking of the global $U(1)$ symmetry comes associated with a massless Goldstone boson, which mediates a long-range interaction among the strings and keeps their energy density finite by effectively introducing a radial cut-off.~\footnote{Note that in a global $U(1)$ scenario like the one under consideration, the energy density of an infinite and isolated string should be expected to diverge since there is no gauge field to compensate the variation of the phase at large distances \cite{Hindmarsh:1994re}.}
The width of the cosmic strings is inversely proportional to the expectation value of the $\chi$ field, which, as shown in Eq~\eqref{eq:chi_roll},  rapidly increases during the first stages of kinetic domination. Since the Hubble rate is the only relevant scale in our model, we have ``fat'' strings with a width $\sim H^{-1}$ soon after the spontaneous symmetry breaking.

Fat strings like the ones under consideration have been extensively studied in the context of axion dark matter \cite{Moore:2016itg,Gorghetto:2018myk,Vaquero:2018tib}, finding an average of one string per Hubble volume. In our case, this is even more expected since the  width of the string is already proportional to the Hubble rate at the time of formation.  Note that this unusual property restricts the formation of cosmic string loops, since the length of these objects cannot be --- for obvious reasons --- shorter than its width. This allows us to neglect the (usually dominant) GW production by loops \cite{Damour:2001bk,Siemens:2006yp,Kawasaki:2011dp} and estimate the total energy density of the network by computing the energy density stored in a single string of Hubble length  $L\propto H^{-1}$ within a Hubble volume $H^3$, namely\footnote{
For simplicity, we disregard the time dependence that logarithmic corrections to the energy density may induce \cite{Moore:2016itg}. The estimates presented in this paper are expected to be robust upon the inclusion of this subleading effect.}
\be
 \rho_{\rm cs} \approx \mu(t) \times L \times H^{3} \approx \mu(t) H^2\,,
\ee
with
\be
 \mu(t) \sim 
 \vert{\chi}(t)\vert^2\,,
\ee
the energy of the string per unit length.\footnote{We omit here a numerical prefactor depending logarithmically on the ratio between the width and the radial cut-off of the string \cite{Hindmarsh:1994re,Moore:2016itg}.}
Note that the relative energy density of the strings as compared to the background component depends on time,
\be\label{noscaling}
\Omega_{\rm cs}\equiv \frac{\rho_{\rm cs}}{3H^2M_{\rm P}^2}\sim\frac{\mu(t)}{M_{\rm P}^2}
\sim \frac{\vert \chi(t)\vert^2}{M^2_{\rm P}}\,,
\ee
meaning that an exact scaling regime \cite{Bennett:1989ak,Allen:1990tv,BlancoPillado:2011dq,Ringeval:2005kr} is never achieved.  This is not necessarily a problem as long as the energy density  of cosmic strings stays subdominant with respect to the background component, i.e. $\Omega_{\rm cs}\ll 1$. This consistency condition translates into a limit on the maximum value that can be explored by the $U(1)$ field, restricting it be sub-Planckian,
\begin{equation}\label{cons_cond}
\vert \chi(t_{\rm tr})\vert\ll M_P\,.
\end{equation}
and setting a maximum on the number of e-folds \eqref{eq:DN} required to arrive to the time-dependent minimum \eqref{eq:chimint},
\begin{equation}\label{eq:DN_crit}
\Delta N_{\rm tr} \ll \frac{1}{\sqrt{6\xi}}\left[18+\ln\left(\frac{1}{1+\sqrt{6\xi}}\right)+\ln\left(\frac{10^{11}\,{\rm GeV}}{H_{\rm kin}}\right)\right]\,.   
\end{equation}

In the presence of a scaling regime, the evolution of the cosmic strings' energy momentum tensor in the vicinity of the horizon  leads to the production of  a scale-invariant gravitational wave spectrum \cite{Krauss:1991qu} (for a recent review see Ref.~\cite{Caprini:2018mtu}). Note, however, that the vacuum expectation value of the scalar field in our scenario evolves with time. We expect therefore a residual scale dependence in the spectrum, measured exactly by the change of this quantity. The amount of GWs produced by a vibrating string of length $L\propto H^{-1}$ can be estimated by approximating the associated quadrupole $Q$ by that of a cylinder of mass $M\approx \mu\, L$ and width $R\propto H^{-1}$ \cite{Krauss:1991qu,Maggiore:1999vm}, namely 
\be
 Q(t) \propto MR^2 \sim \mu(t) H^{-3}\,,
\ee
with the precise proportionality constant depending on the level of non-sphericity of the configuration.
Assuming that the relaxation time is of the order of $H$, the luminosity in GWs becomes
\be\label{eq:luminosity}
 \mathcal{L}(t) \sim M_{\rm P}^{-2}\, \dddot{Q}(t)^2 \sim M_{\rm P}^{-2} (H^3(t) Q(t))^2\sim \mu^2(t)M_{\rm P}^{-2}\,,
\ee
where we have replaced time derivatives in favor of the Hubble parameter $H$. At a given instant $t$, the emission of GWs is peaked around the characteristic scale of the system at that time, decaying as a power-law for  shorter and longer distances. The relative power emitted by the string network in a Hubble time (i.e. per $dN\equiv Hdt$) is given by \cite{Kamada:2015iga}
\be\label{k_dep}
 \frac{\Delta P_{\rm GW}(t,k)}{\Delta N}\equiv\frac{1}{3H^2M_{\rm P}^2}\frac{d\rho_{\rm GW}(t,k)}{d\log k}
 \simeq 
\begin{cases}
P_{\rm peak}(t)
 \left(\frac{k}{k_{\rm peak}(t)} \right)^{\alpha} \hspace{7mm}{\rm for}\hspace{5mm}  k \lesssim k_{\rm peak}\,, \\ 
P_{\rm peak}(t)
 \left(\frac{k}{k_{\rm peak}(t)} \right)^{-\bar\alpha} \hspace{5mm}{\rm for}\hspace{5mm}  k > k_{\rm peak}\,, \\ 
\end{cases}
\ee
with $k_{\rm peak}(t)\sim aH$ the peak emission wavenumber\footnote{We omit here a (larger than one)  proportionality constant ensuring causality \cite{Kamada:2015iga}.} and 
\bea\label{eq:P_peak}
P_{\rm peak}(t)   
 &\sim& 
 \frac{L^{-2}\times\mathcal{L}}{H^2 M_{\rm P}^2}  \sim  \frac{\mathcal{L}}{M_{\rm P}^2}
 \sim 
 \left( \frac{\vert \chi(t)\vert}{M_{\rm P}} \right)^4 \,.
\eea 
The exponents $\alpha$ and $\bar \alpha$ in Eq.~\eqref{k_dep} parametrize our ignorance about the exact momentum dependence. While a scaling $\alpha=2$  can be inferred from simple causality arguments~\footnote{The modes with $ k \lesssim k_{\rm peak}$ at a given time correspond to super-horizon scales.} \cite{Dufaux:2007pt,Kamada:2015iga}, the value of $\bar \alpha$ must be derived from simulations. Note, however, that this exponent is intimately related to the dynamics of small scales and it should be therefore rather independent of the specific expansion history. This observation allows us to benefit from existing numerical results to set a fiducial value $\bar\alpha\sim 2$ ~\cite{Kamada:2015iga}.
 
\subsection{GW spectrum in a generic background}

The precise form of the accumulated GW spectrum at large momenta is obtained by integrating Eq.~\eqref{k_dep} from the time $t_{\rm i}$ at which the strings start to form to an arbitrary time $t$ that will be later associated with the moment at which strings decay and the production of GWs is halted. Since the emission peak in our scenario is set by the comoving horizon, $k_{\rm peak} \sim a H=\tau^{-1}$, it is convenient to work in conformal time $\tau$. In terms of this time coordinate, the integrated GW spectrum takes the form
\bea \label{eq:OmegaGW-gen}
 && \Omega_{\rm GW}\left( \tau,k \right) 
 =\int^{\tau}_{\tau_{\rm i}} d \log \tau' 
 \frac{\Delta P_{\rm GW} (\tau',k)}{ \Delta \log \tau' }\left(\frac{a(\tau')}{a(\tau)} \right)^b \,,
 \eea 
 where we have used the fact that $dN\propto d\log\tau$ and accounted for the evolution of the GW energy density. For the sake of completeness and in order to facilitate the comparison with other works in the literature, 
we have considered a generic power-law evolution $a\sim t^p\sim \tau^{p/(1-p)}$ with $p$ a constant.~\footnote{The choice of  constant $p$ is indeed a very good approximation \textit{within a given cosmological epoch}. Although the formalism presented here could be easily extended to account for smooth transitions among different cosmological eras by simply promoting $p$ to $p(\tau)$, this will not be necessary since we will focus here on the GW production during kinetic domination.}
The scaling power $b$ in Eq.~\eqref{eq:OmegaGW-gen} is related to $p$ via $b\equiv 4-2/p$ and takes values $b=-2$ and $b=0$ during kination and radiation domination.

We can distinguish two possible scenarios depending on the relation between the bare mass $m_\chi$ and the Hubble rate $H_{\rm rad}$ at the time of heating: 
\begin{enumerate}
\item If $m_\chi \geq H_{\rm rad}$ the symmetry is restored before radiation domination and the $U(1)$ scalar field is free to oscillate around the origin, leading to the evanescence of the cosmic string network and halting the GW production. 
\item If $m_\chi<H_{\rm rad}$ the $U(1)$ field remains frozen at its vacuum expectation value after symmetry restoration. Although additional strings are no longer generated, the existing ones will continue producing GWs after reentering the expanding horizon. This GW production will stop when the instantaneous Hubble rate becomes comparable to the mass of the field, $H\lesssim m_\chi$. From there on, the field $\chi$ will start oscillating around the origin and the strings will consequently decay. 
\end{enumerate} 
Since the use of numerical simulations is most probably required in the latest scenario, we will focus here on the $m_\chi \geq H_{\rm rad}$ case.
The total power in GWs produced by the cosmic string network can be then computed by integrating Eq.~\eqref{eq:OmegaGW-gen} from the formation time $\tau_i$ to the time $\tau = \tau_{\rm osc}$ at which the symmetry is restored. Taking into account Eq.~\eqref{k_dep} we can rewrite this expression as \cite{Kamada:2015iga}
\begin{eqnarray}\label{eq:integral_split}\nonumber
\Omega_{\rm GW}\left( \tau_{\rm osc},k \right)  &\simeq&
 \int^{\tau_k}_{\tau_{\rm i}} \frac{d\tau' }{\tau'}P_{\rm peak}(\tau') 
\left( \frac{k}{k_{\rm peak}(\tau')} \right)^{\alpha }\left(\frac{a(\tau')}{a(\tau)} \right)^b \\  &+& 
 \int^{\tau_{\rm osc}}_{\tau_k} \frac{d\tau' }{\tau'}P_{\rm peak}(\tau')  
\left( \frac{k}{k_{\rm peak}(\tau')} \right)^{-\bar \alpha } \left(\frac{a(\tau')}{a(\tau)} \right)^b \,,
\end{eqnarray}
where we have defined $\tau_k$ as the time at which the peak emission mode coincides with the mode $k$, i.e. $k_{\rm peak}(\tau_k)=k$. This splitting takes into account the amount of time that each mode spends in the blue or red part of the spectrum given that the peak is evolving with time. Also notice that, since we require that $\tau_k>\tau$, the integral is only valid for $k>k_{\rm peak}(\tau)$ \cite{Kamada:2015iga}. 

The integrated GW spectrum in Eq.~\eqref{eq:integral_split} has three main contributions. On the one hand, we have two associated with the blue-shifting of the power spectrum 
itself, which give either a $k^{-\bar \alpha}$ or a $k^\alpha$ scaling depending on how much time did the wavenumber $k$ spend in each slope.
On the other hand, we have the contribution coming from the time at which the mode $k$ coincides with the generic time-dependent peak emission mode $k_{\rm peak}(\tau_k)$. In that case the power spectrum is evaluated at $k\,\tau_k\sim1$. Assuming a generic power-law dependence  $P_{\rm peak}\sim \tau^{-\gamma}$ with constant parameter  $\gamma$,~\footnote{As we will see in the next section, the precise value of $\gamma$ depends on the $U(1)$ field dynamics and on the model parameters. Nevertheless, for an easier comparison with the literature, we report here that for the field evolution in  Eq.~\eqref{eq:chimint}, we have $\gamma= -\frac{8}{(n-2)(1-p)}$. Using this expression, we can easily recover the results of Ref.~\cite{Kamada:2014qja} by simply setting $p=2/3$ and $n\geq 6$, cf.~also Section \ref{sec:GW_today}.  } this leads to a scaling $\Omega_{\rm GW}\sim k^\beta$ with 
\begin{equation}\label{eq:beta_gamma}
\beta\equiv \gamma -\frac{4\, p-2}{1-p}\,.
\end{equation}
The above contributions compete in the integrated form of Eq.~\eqref{eq:integral_split} yielding~\footnote{Notice that for $\beta-\alpha=0$, the spectrum has a different solution, namely
\begin{equation}
\Omega_{\rm GW}(\tau_{\rm osc},k)\simeq P_{\rm peak}(\tau_{\rm osc})\left(\frac{k}{k_{\rm peak}(\tau_{\rm osc})}\right)^\beta\left[\log\left(\frac{k_{\rm peak}(\tau_{\rm i})}{k}\right)+\frac{1}{\beta+\bar \alpha}\left(1-\left(\frac{k_{\rm peak}(\tau_{\rm osc})}{k}\right)^{\bar\alpha+\beta}\right)\right]\,.\nonumber
\end{equation}
Hence, its power law behavior for intermediate frequencies is $\Omega_{\rm GW}(\tau_{\rm osc},k)\propto k^\beta \log(k_{\rm peak}(\tau_{\rm i})/k)$ rather than just $k^\beta$. A similar logarithmic dependence can be found in the $\bar \alpha+\beta=0$ case.}
 \bea \label{eq:OmegaGW_early}\nonumber
\Omega_{\rm GW}\left( \tau_{\rm osc},k \right)  &\simeq &
P_{\rm peak}(\tau_{\rm osc}) 
\left( \frac{k}{k_{\rm peak} (\tau_{\rm osc})} \right)^{\beta}\times\\
&\times&\left[ \frac{1}{\beta- \alpha}\left( 
\left( \frac{k_{\rm peak} (\tau_{\rm i})}{k}\right)^{\beta-\alpha}  -1\right)
\frac{1}{\bar \alpha + \beta} 
\left(1 -\left( \frac{k_{\rm peak} (\tau_{\rm osc})}{k} \right)^{\bar\alpha  + \beta} \right)\right]\,.
\eea 
This analytical expression explores a range of frequencies comprised between $k_{\rm peak}(\tau_{\rm osc})\le k \le k_{\rm peak}(\tau_{\rm i})$. For intermediate frequencies well within the integration limits, $k_{\rm peak}(\tau_{\rm osc})\ll k \ll k_{\rm peak}(\tau_{\rm i})$, we can identify three regimes depending on the sign of $\beta-\alpha$ and $\beta+\bar\alpha$, namely 
\be\label{eq:power_index}
  \Omega_{\rm GW}
 \propto 
\begin{cases}
k^\alpha \hspace{7mm}{\rm for}\hspace{5mm}  \beta-\alpha>0\,, \\ 
k^\beta \hspace{7mm}{\rm for}\hspace{5mm}  \beta-\alpha<0 \quad \text{and}\quad  \bar\alpha+\beta>0\,, \\ 
k^{-\bar\alpha} \hspace{5mm}{\rm for}\hspace{5mm}  \beta-\alpha<0 \quad \text{and}\quad  \bar\alpha+\beta<0\,. \\
\end{cases}
\ee
Since $\alpha>0$, the first case corresponds to a blue-tilted spectrum while the second one can be either blue or red depending on the sign of $\beta$, which is ultimately determined by the evolution of the $\chi $ field and the background, cf. Eq.~\eqref{eq:beta_gamma}.  Finally, the last case corresponds to a red tilted spectrum since $\bar\alpha>0$ \cite{Kamada:2015iga}.

\subsection{GW spectrum in quintessential inflation}\label{sec:Int_spec_Kin}

The formalism presented in the previous section is quite general and holds for any background and temporal dependence of the emission peak, as long as this can be parametrized by a power law  $P_{\rm peak}\sim \tau^{-\gamma}$ with $\gamma$ approximately constant for some temporal interval $\Delta\log\tau$. In this section, we particularize our results to the kinetic-dominated era appearing in our quintessential inflation scenario, determining the precise form of $\gamma$ and the power spectrum as a function of the model parameters. 

The background evolution during kinetic domination is specified by taking $p=1/3$ and hence $b=-2$. During this epoch, the dynamics of the scalar field $\chi$ can be split into a \textit{rolling} and a \textit{minimum phase}, each of them leading to different characteristic shapes in the GWs spectrum. The first phase is associated with the initial stages of kination, where the $U(1)$ scalar field is still rolling down the origin towards the Hubble-induced minimum of the potential. The second phase describes the tracking of the minimum once the field reaches it, if it does. We refer the interested reader to the Appendix \ref{sec:ais}, where a detailed calculation of the GW spectrum is performed.\footnote{It should noted that we are implicitly assuming that the cosmic string network forms and starts to emit gravitational waves right after the symmetry breaking. To the best of our knowledge, such transition has not been studied in numerical simulations but we expect  our order of magnitude estimates to stay valid.} Here we will simply discuss the shape of the spectrum and its dependence on the parameters.
\begin{table}
\begin{center}
\begin{tabular}{lccccccc}    \hline\hline
Phase & \hspace{2mm}  $\alpha$ & $\hspace{2mm}\bar \alpha$ & \hspace{2mm} $\beta$ &  $\hspace{5mm}\gamma$  & \hspace{5mm}$\beta-\alpha$ &$\bar\alpha+\beta$& \hspace{2mm}  tilt range \Tstrut\Bstrut \\
 \hline
\textit{Rolling} & \hspace{2mm}  2 & $\hspace{2mm}\sim 2$
 & \hspace{2mm} $1-2\sqrt{6 \xi}$ & $\hspace{2mm}-2\sqrt{6 \xi}$  &\hspace{2mm} $-1-2\sqrt{6 \xi}$& \hspace{2mm} $3-2\sqrt{6 \xi}$ & \hspace{2mm}  $1$ to $-2$ \Tstrut\Bstrut\\ 
 
\textit{Minimum} &\hspace{2mm}   2 & $\hspace{2mm} \sim 2$& \hspace{2mm} $\dfrac{10+n}{n-2}$ &  \hspace{2mm} $\dfrac{12}{n-2}$ & \hspace{2mm} $\dfrac{14-n}{n-2}$ & \hspace{2mm} $3\dfrac{n+2}{n-2}$ & \hspace{2mm} $1$ to $2$ \Tstrut\Bstrut \\   
\hline\hline 
\end{tabular}
\end{center}
 \caption{Values of the spectrum powers during the \textit{rolling} and the \textit{minimum phase}. The value of $\gamma$ has been extracted from Eqs.~\eqref{eq:chimint}, \eqref{eq:chi_roll} into \eqref{eq:P_peak}. The tilt range has been computed using Eq.~\eqref{eq:power_index} taking the \textit{formal} limits $\xi\rightarrow 0$, $\xi\rightarrow \infty$ and $n\rightarrow 4$, $n\rightarrow \infty$, respectively.}\label{tab:table2}
\end{table}
  In particular, we report in Table \ref{tab:table2} the values of the power law indexes during the various stages. Note that the parameter $\gamma$ is negative during the \textit{rolling phase} $(\tau<\tau_{\rm tr})$ and positive definite during the \textit{minimum phase} $(\tau>\tau_{\rm tr})$. As shown in the last column, the spectrum of GWs  during the \textit{rolling phase} can be both blue and red depending on the values of the parameter $\xi$, namely
\begin{align}\label{eq:Om_scaling_roll}
\Omega_{\rm GW}(\tau,k>k_{\rm tr})\propto
\begin{cases}
k^{1+\gamma} \hspace{5mm}{\rm if}
\hspace{5mm} \xi<3/8\,, \\ 
k^{-\bar\alpha} \hspace{7mm}{\rm if}
\hspace{5mm} \xi>3/8 \,.
\end{cases}
\end{align}
On the other hand, the spectrum  during the \textit{minimum phase} can only be blue, with the precise tilt depending on the order of the higher-order operators in Eq.~\eqref{LBL},
 \begin{align}\label{eq:Om_scaling_min}
\Omega_{\rm GW}(\tau,k<k_{\rm tr})\propto
\begin{cases}
k^{1+\gamma} \hspace{7mm}{\rm if }\hspace{5mm} n>14\,, \\ 
k^{\alpha}   \hspace{10mm}{\rm if} \hspace{5mm}  n<14 \,.
\end{cases}
\end{align}
The integrated GW spectrum when both phases are present can be generically written as an amplitude times a $k$-dependent form function, namely
\begin{equation}\label{eq:spectrum_formal}
 \Omega_{\rm GW}(\tau_{\rm osc},k)\sim P_{\rm peak}(\tau_{\rm kin})\Theta^{-1}\left(\frac{a_{\rm osc}}{a_{\rm rad}}\right)^2 F\left(\kappa,s_{\rm osc},s_{\rm tr}\right)\,,
\end{equation}
where the blue-shifting factor $a_{\rm osc}/a_{\rm rad}$ appears because we factored out $\Theta^{-1}$ and we have defined the dimensionless variables
\begin{equation}\label{eq:s_and_x}
 s\equiv \frac{\tau}{\tau_{\rm kin}}\hspace{12mm} \text{and}\hspace{12mm}\kappa\equiv \frac{k}{a_{\rm kin}H_{\rm kin}}\,.
\end{equation}
The functional $F$ encodes the momentum behavior of the spectrum as a function of the model parameters. A detailed computation of this quantity assuming  a sudden transition between the \textit{rolling} and the \textit{minimum phases} can be found in the Appendix \ref{sec:ais}.\footnote{Note that this slightly overestimates the size of spectrum since the transition from the \textit{rolling} to the \textit{minimum phase} is expected to be smooth rather than instantaneous.} 

\begin{figure}
\begin{center}
\includegraphics[scale=0.9]{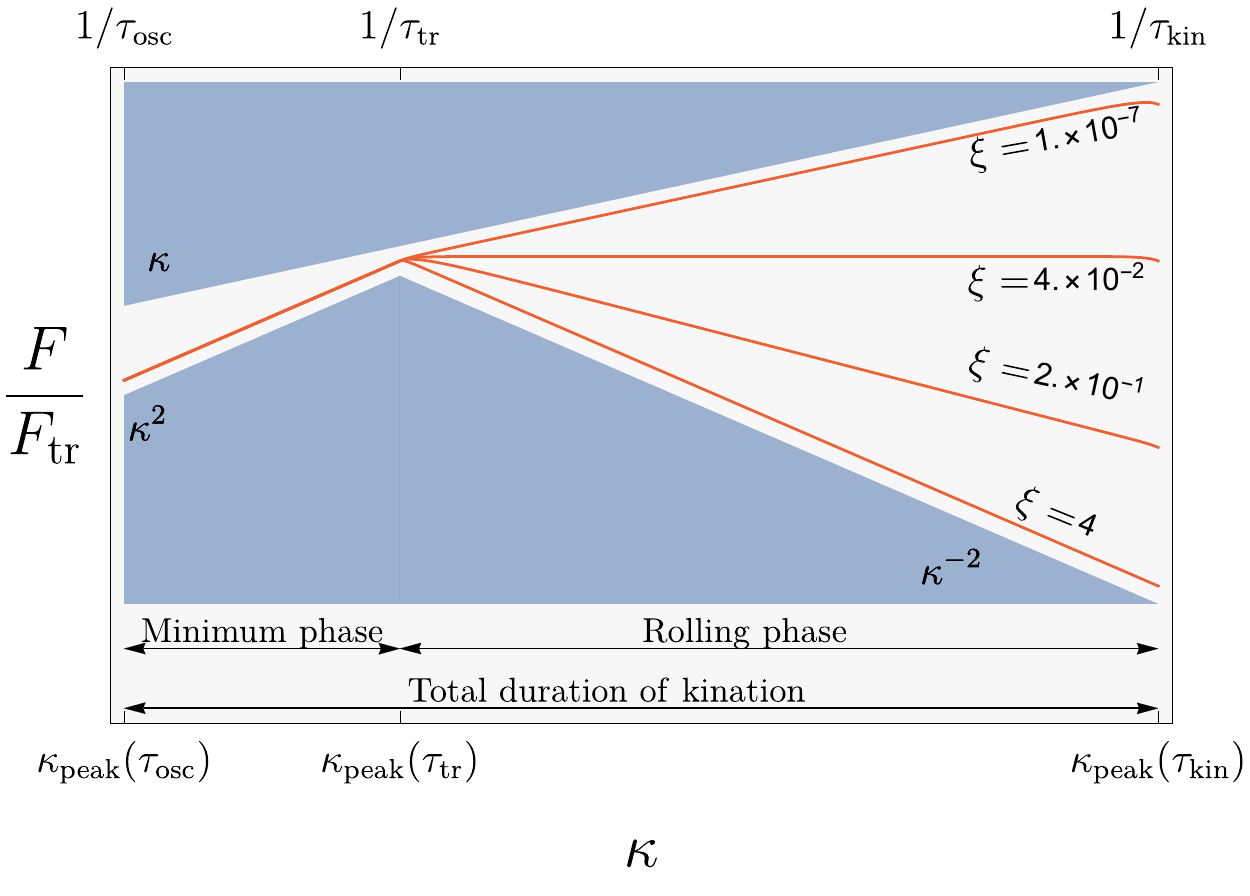}
\caption{Form of the function $F$ in Eq.~\eqref{eq:spectrum_formal} for various values of $\xi$. The shaded area represents the asymptotic values reported in Table~\ref{tab:table2}. Here we have assumed $a_{\rm osc}=a_{\rm rad}$ and $n=4$.}
\label{fig:F_vs_x}
\end{center}
\end{figure}
As illustrated in Figure~\ref{fig:F_vs_x}, the shape of the spectrum is determined by the values of the $\alpha$, $\bar \alpha$ and $\beta$ parameters, according to the discussion in Eq.~\eqref{eq:power_index}. In particular, the two stages of evolution give rise to two spectral indexes, hence to an inflection point in the spectrum. 
We can distinguish three characteristic $k$ values, namely $k_j\equiv a(\tau_j) H(\tau_j)$ with $j={\rm kin},{\rm tr},{\rm rad}$ standing respectively for the symmetry breaking time, the onset of the \textit{minimum phase} and that of radiation domination.\footnote{
We could additionally estimate what would happen if the universe enter the radiation domination regime before the $U(1)$ field decays. Assuming that the field stays frozen at the value of restoration for a while -- i.e. $P_{\rm peak}\approx {\rm constant}$ -- and taking into account that the background energy density scales in that case as the one in GWs, the only contribution will be that from $k\,\tau_k\sim 1$, leading then to a scale-invariant spectrum. If that were the case, there would be a plateau for small $k<k_{\rm rad}$ in the power spectrum with a cut-off when the field decays. We will not further pursue this possibility in this paper, but it is worth noting the possibility.} 
For each of these modes the amplitude of the power spectrum can be determined as follows: using Eq.~\eqref{eq:F_min} one can get the amplitude for $k=k_{\rm kin}$ and then use the scaling in Eqs.~\eqref{eq:Om_scaling_roll} and \eqref{eq:Om_scaling_min} to compute the amplitude for the other two modes. Assuming for simplicity  that both phases are present and that  they have a duration of several e-folds, i.e.  $k_{\rm rad}\ll k_{\rm tr}\ll k_{\rm kin}$, i.e.  
we get
\begin{align}\label{eq:Omega_amplitude_B}
\hspace{-1mm}\Omega_{\rm GW}(\tau_{\rm osc},k_{\rm kin}) \propto 
\left(\frac{\vert \chi_{\rm min}(a_{\rm kin})\vert}{M_{\rm P}}\right)^4 \left(\frac{a_{\rm osc}}{a_{\rm rad}}\right)^2 \Theta^{-1}
\begin{cases}
1 \hspace{30mm}{\rm if}
\hspace{3mm} \bar\alpha +\beta>0\,, \\ 
 \left(\frac{a_{\rm rad}}{a_{\rm tr}}\right)^{2(\bar\alpha+\beta)}\Theta^{\bar \alpha+\beta} \hspace{3mm}{\rm if}
\hspace{3mm} \bar\alpha+\beta<0 \,,
\end{cases}
\end{align}
and 
\begin{align}
\label{eq:Omega_amplitude_eq}
\Omega_{\rm GW}(\tau_{\rm osc},k_{\rm tr})&=\Omega_{\rm GW}(k_{\rm kin})\left(\frac{a_{\rm kin}H_{\rm kin}}{a_{\rm tr} H_{\rm tr}}\right)^{c}\,,
\\
\label{eq:Omega_amplitude_R}
\Omega_{\rm GW}(\tau_{\rm osc},k_{\rm rad})&= \,\Omega_{\rm GW}(k_{\rm tr})\left(\frac{a_{\rm rad}H_{\rm rad}}{a_{\rm tr} H_{\rm tr}}\right)^{d}\,,
\end{align}
where in the $\bar \alpha+\beta<0$ case there is an extra $a_{\rm tr}/a_{\rm rad}$ term because we factored out $\Theta$ and we have defined exponents $c=\rm{max}(-\bar \alpha, 1+\gamma)$ and $d=\rm{max}(\alpha, 1+\gamma)$.  Note that the heating efficiency governs the range of modes over which the spectrum extends,
\begin{equation}\label{eq:delta_k}
\frac{k_{\rm  peak}(\tau_{\rm kin})}{k_{\rm peak}(\tau_{\rm osc})} = \Theta^{-1}\left(\frac{a_{\rm osc}}{a_{\rm rad}}\right)^2\,,
\end{equation}
 becoming smaller for increasing $\Theta$.

\section{GW spectrum today}\label{sec:GW_today}

Once the symmetry is restored, the strings decay and the production of GWs terminates. Since the energy density of the residual GW background scales as radiation, the GWs spectrum at the present cosmological epoch can be well approximated by~\footnote{ In deriving this expression we have neglected the variation of the effective number of degrees of freedom. This effect can be easily incorporated by replacing
\begin{equation}
 \Omega_{\rm GW} ( \tau_{\rm rad})\,\,\,\rightarrow\,\,\, \left( \frac{g_s (\tau_0)}{g_s(\tau_{\rm rad})} \right)^{4/3} 
 \left( \frac{g_*(\tau_{\rm rad})}{g_* (\tau_0)} \right) 
 \Omega_{\rm GW} ( \tau_{\rm rad}) \,, \nonumber
\end{equation} 
with $g_s$ and $g_*$ the entropic and relativistic degrees of freedom at a given time/temperature. Note, however, that, for the order of magnitude estimates presented in this paper, the additional factors play almost no role.~Indeed, for $g_s\sim g_*$ the correction to Eq.~\eqref{Omeganow} is  ${\cal O}(1)$.}  
\bea\label{Omeganow}
\Omega_{\rm GW} (\tau_0,k)h^2 &=&  \Omega_{\rm R} h^2 
\left( \frac{a_{\rm rad}}{a_{\rm osc}}  \right)^{2
 } \,
 \Omega_{\rm GW} ( \tau_{\rm osc},k)\,,
\eea
with $\tau_0$ the current conformal time, $\Omega_{\rm R}  h^{2}\simeq 4.15 \times 10^{-5}$ the radiation energy density nowadays and $h=0.678$. The momentum dependence of this expression can be made explicit by combining it with  Eq.~\eqref{eq:spectrum_formal},
\begin{equation}\label{Omeganowk}
 \Omega_{\rm GW} (\tau_0,k)h^2 = \Omega_{\rm R} h^2  
  P_{\rm peak}(\tau_{\rm kin})\Theta^{-1}F(\kappa,s_{\rm osc},s_{\rm tr})\,.
\end{equation}
In order to connect the mode interval in Eq.~\eqref{Omeganowk} with the present-day frequency range, we make use of the standard relation among frequencies and wavenumbers, namely
\begin{equation}
 f_0=\frac{1}{2 \pi}\frac{k}{a_0}\,,   
\end{equation}
with the subscript $0$ referring to present day quantities. Taking into account the cosmological evolution during radiation domination, we can rewrite this expression as 
\begin{equation}\label{eq:x_to_f}
\kappa=\frac{2 \pi f_0}{\sqrt{H_0 H_{\rm kin}}}\Theta^{1/4}\,,
\end{equation}
with $\kappa$ the dimensionless variable in Eq.~\eqref{eq:s_and_x} and  $H_0\simeq 100\, h \,{\rm km}/{\rm s\, Mpc}^{-1}$ the present Hubble rate.
Combining this result with 
Eq.~\eqref{eq:delta_k} we find that the range of frequencies generated by our scenario lays between
\be 
f_0^{\rm kin}\sim 3\times 10^{11} \,{\rm Hz} \left(\frac{H_{\rm kin}}{10^{11} {\rm GeV}}\right)^{1/2}\, \left(\frac{\Theta}{10^{-14}}\right)^{-1/4}\,,
\ee
and
\be
f_0^{\rm osc}\sim 3\times 10^{-3} \,{\rm Hz} \left(\frac{H_{\rm kin}}{10^{11} {\rm GeV}}\right)^{1/2}\, \left(\frac{\Theta}{10^{-14}}\right)^{3/4}\left(\frac{a_{\rm rad}}{a_{\rm osc}}\right)^2 \,,
\ee
with the superscripts kin and rad referring respectively to the evaluation of the spectrum at $k=k_{\rm peak}(\tau_{\rm kin})$ and $k=k_{\rm peak}(\tau_{\rm osc})$. Within this frequency interval, we can distinguish the transition between the \textit{rolling phase} and the \textit{minimum phase} (cf.~Section \ref{sec:Int_spec_Kin}), namely
\begin{equation}
 f_0^{\rm tr}\sim 0.2 \,{\rm Hz} \left(\frac{H_{\rm kin}}{10^{11} {\rm GeV}}\right)^{1/2}\, \left(\frac{\Theta}{10^{-14}}\right)^{-1/4}  \exp\left(\frac{\Delta N_{\rm tr}}{14}\right)\,,
\end{equation}
with the superscript tr denoting the evaluation of the spectrum at $k=k_{\rm peak}(\tau_{\rm tr})$ and $\Delta N_{\rm tr}$ given by Eq.~\eqref{eq:DN}.

\begin{figure}
\begin{center}
\includegraphics[scale=0.7]{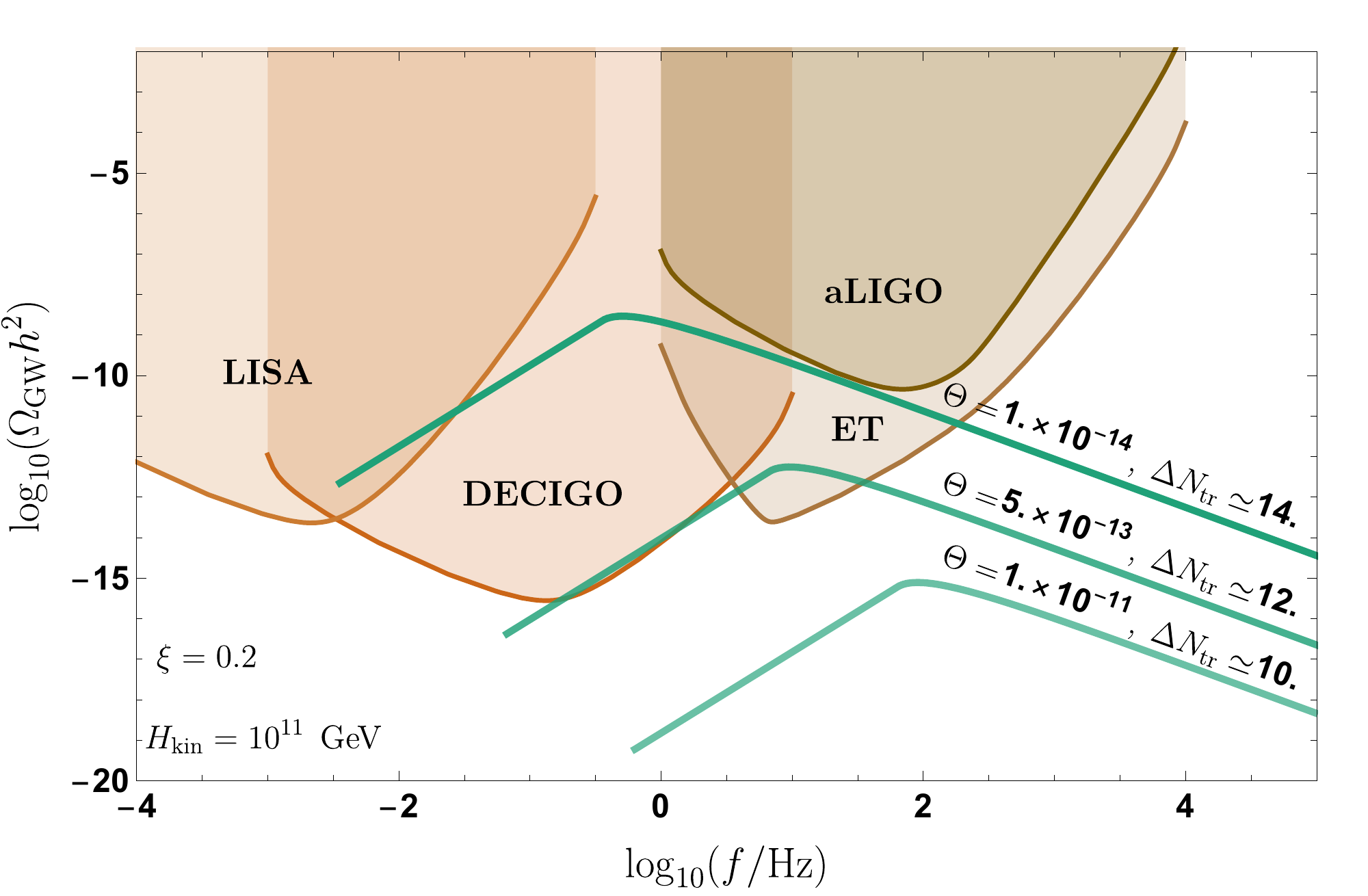}
\caption{Comparison among the GW spectrum produced by the global cosmic string network during kinetic domination and the  power-law integrated sensitivity curves of various GW experiments \cite{Thrane:2013oya,Breitbach:2018ddu}.
Here we have assumed that the strings are immediately destroyed at the onset of radiation domination ($a_{\rm osc}=a_{\rm rad}$) and chosen fiducial values $H_{\rm kin}=10^{11}$ GeV and $\xi = 0.2$. For this choice of parameters the backreaction of the cosmic string network is always small and the big bang nucleosynthesis constraint \eqref{GWbound1} is satisfied.
}
\label{fig:Om_vs_f2}
\end{center}
\end{figure}

Ground-based interferometers such as LIGO \cite{adv_ligo}, VIRGO \cite{virgo} or the Einstein Telescope \cite{et} are already sensitive to frequencies around $f\sim 100$ Hz, while space interferometers such as eLISA \cite{lisa} or DECIGO \cite{decigo} will be able to cover a $0.1$ mHz-$10$ Hz frequency window. The GW spectrum predicted by our mechanism is confronted with the  power-law integrated sensitivity curves\footnote{ These curves take into account the enhancement in detector sensitivity following the integration over frequency on top of the integration over time.} 
of these experiments
in Fig.~\ref{fig:Om_vs_f2}. As shown in this plot, a potentially observable window within the scope of  future surveys can be easily found for not too large values of $H_{\rm kin}$ and relatively small heating efficiencies. 
It is important to notice that the produced spectrum is not restricted by CMB \textit{anisotropies}. In particular, the production of GWs by Hubble-induced cosmic strings happens i) well-inside the horizon, ii) has a limited duration and iii) is restricted to highly energetic frequencies significantly exceeding those associated to the Hubble  scale at decoupling, namely  $3.4\times 10^{-19} \,{\rm Hz}<f_{\rm CMB} < 2.1\times 10^{-17}\, {\rm Hz}$ \cite{Caprini:2018mtu}. The only restriction comes from its (integrated) contribution to the \textit{background} energy density in Eq.~\eqref{GWbound1}. Note also that although our results are based on pure dimensional analysis and could therefore differ from the actual values by a few orders of magnitude (see for instance Ref.~\cite{Figueroa:2012kw}) such uncertainty would not necessarily affect the observability of the model. In particular, a potential change in the amplitude of the spectrum could be easily compensated by a change on the number of e-folds spent in the \textit{rolling phase} and on the duration of the kinetic dominated era.

It is interesting to compare our findings with previous results in the literature.  On general grounds, our mechanism makes use of two basic ingredients: a non-minimal coupling to gravity\footnote{The use of non-minimal couplings in the context of topological defects goes back to Ref.~\cite{Yokoyama:1988zza}, where they were introduced to avoid the generation of cosmic strings during inflation. Note, however, that the post-inflationary cosmic string formation described in this reference happens due to an explicit symmetry-breaking potential and not to the Ricci scalar itself.} and a period of kinetic domination. The fact that we are dealing with a Hubble-induced symmetry breaking pattern has far reaching consequences. First, the width of the strings in our scenario is proportional to the Hubble rate at the time of formation. Second, the temporal dependence in Eq.~\eqref{noscaling} forbids the appearance of an exact scaling regime. Third, the eventual restoration of the symmetry at the onset of radiation domination rescues the model from the limitations associated to (long-lived) non-scaling topological defects. All these properties are genuinely new, marking significant differences with the usual global string scenarios \cite{Hagmann:1990mj,Yamaguchi:1998gx,Yamaguchi:1999yp,Yamaguchi:1999dy}. This applies also to other studies in the literature involving also a kinetic dominated era \cite{Cui:2017ufi,Cui:2018rwi,Artymowski:2017pua}. Interestingly, a restoration mechanism similar to the one presented in this paper was used in  Ref.~\cite{Kamada:2015iga} in a matter dominated era.
Although the shape of the spectrum in the latest scenario shares some similarities with that in Fig.~\ref{fig:Om_vs_f2}, the associated physics is completely different, in particular: 
\begin{enumerate}[i)]
\item The energy density parameter $\Omega_{\rm GW}$  grows during kinetic domination rather than decreasing.
\item The knee in the spectrum in our scenario is not due to the transition from matter to radiation domination, but rather to the evolution of the $U(1)$ field from a \textit{rolling phase} in which the energy density of the strings increases to a \textit{minimum phase} in which the energy density of the strings decreases. 
\item   The spectral tilt at large frequencies depends in our case on  the non-minimal coupling to gravity, rather than on the dimension of higher-dimensional operators. 
\item Contrary to what happens in matter domination, the low frequency band of the spectrum is not always proportional to $k^{2}$, but rather depends on the detailed structure of the scalar potential. 
\end{enumerate}

Although the existence of a \textit{rolling phase} is crucial for the GWs produced in our scenario to be within the observable band, its absence is also interesting by itself.   In particular, the spectrum generated by an the \textit{minimum phase} alone is \textit{always} blue during kinetic domination, meaning that the leading contribution at the dominant high-frequency end $k_{\rm kin}$ can be easily constrained by the integrated bound in Eq.~\eqref{GWbound1}, namely
\begin{align}
\Omega_{\rm GW}(k_{\rm kin})\sim 10^{-5}\,\Theta^{-1}\left(\frac{\vert \chi_{\rm min}(a_{\rm kin})\vert}{M_{\rm P}}\right)^4 < 10^{-6}\,.
\end{align}
This translates into a strong restriction on the presence of non-minimal couplings and on the duration of a kinetic dominated stage. Either the field stays sub-Planckian or the heating efficiency is high. In both cases, the GWs spectrum would never hit the observational band due to its  strong blue tilt. 

\section{Conclusions}\label{sec:conclusions}

The free propagation of gravitational waves upon production makes them a perfect cosmological probe for retrieving information on energy scales and epochs that are hardly accessible by any other means. Violent processes such as inflation, (re)heating or phase transitions could naturally produce stochastic GW backgrounds within the reach of  present and future GW interferometers.

In this paper we have considered the GW spectrum generated by a network of short-lived cosmic strings appearing in runaway quintessential inflation scenarios displaying a non-minimally coupled $U(1)$ spectator field. In this class of models, the Universe enters a kinetic-dominated regime soon after the end of inflation. During this unusual cosmological epoch the Ricci scalar turns negative,  breaking spontaneously  the $U(1)$ symmetry  and triggering the formation of global cosmic strings. These topological defects emit GWs until the onset of radiation domination, where the Ricci scalar vanishes and the symmetry is restored. At this point, the strings decay, halting the GW production and leaving behind a GW background within the scope of near future observational campaigns. Interestingly, the spectrum displays some characteristic features that might allow to discriminate the proposed scenario from other mechanisms in the literature. On the one hand, its tilts are directly related to the non-minimal coupling to gravity and to the structure of the higher dimensional operators ensuring vacuum stability during kinetic domination. On the other hand, the amplitude at peak position is associated to the transition between 
a \textit{rolling phase} associated with the initial stages of kinetic domination and a \textit{minimum phase} in which the field tracks the minimum of the effective potential. In the event of a detection, these features provide a way of testing the structure of the scalar sector and the existence of non-minimal couplings to gravity, seeding some light on the underlying particle physics theory describing  the early Universe. We have also argued that in the absence of a long \textit{rolling phase}, the spectrum is only blue-tilted, leading to a strong restriction on non-minimal couplings to gravity and the duration of a kinetic dominated epoch.
 
The treatment presented in this paper is rather general and not linked to any particular realization of the quintessential inflationary paradigm or to the details of the entropy production process. Among other possibilities, our scenario could take place, for instance, in $\alpha$-attractor models \cite{Dimopoulos:2017zvq,Dimopoulos:2017tud,Akrami:2017cir,Garcia-Garcia:2018hlc} or quantum gravity frameworks. \cite{Wetterich:2014gaa,Hossain:2014xha,Rubio:2017gty}.
The formalism could be additionally extended to accommodate a richer phenomenology, such as the generation of different \textit{short-lived} topological defects or the creation of primordial black holes \cite{Amendola:2017xhl}. An interesting possibility along this line of thinking would be to identify the $U(1)$ charge with the baryonic number, while allowing for a small and explicit breaking of the $U(1)$ symmetry at large field values \cite{Bettoni:2018utf}. This framework could generate a non-vanishing global baryon asymmetry together with a \textit{correlated} GW background, opening the possibility of testing baryogenesis with GWs.  Determining the precise dynamics of this appealing scenario would most probably require the use of numerical simulations. We postpone the detailed analysis to a future work.

\section*{Acknowledgments}
DB and JR acknowledge support from DFG through the project TRR33 ``The Dark Universe'' during the first stages of this work. GD is partially supported by DFG Collaborative Research center SFB 1225 (ISOQUANT). JR thanks Asier Lopez-Eiguren and David Weir for useful discussions. GD would like to thank A.~Kamada, M.~Sasaki and T.~Sekiguchi for useful discussions.  We finally thank the anonymous referee for useful and constructive comments.

\appendix 
\section{Detailed computation of GW spectrum}\label{sec:ais}

In this appendix we give details on the computation of the integrated spectrum  discussed in Sections \ref{sec:Int_spec_Kin} and \ref{sec:GW_today} and plotted in Figs.~\ref{fig:F_vs_x} and \ref{fig:Om_vs_f2}. Our approach is based on two major approximations. First, we assume a sudden transition between the \textit{rolling} and the \textit{minimum phases}. Secondly, we slightly extrapolate the \textit{rolling phase} solution  beyond its regime of validity.
Note, however, that none of these working assumptions has a significant impact on phenomenology. A more sophisticated analysis would merely provide more accurate restrictions on the model parameters, without significantly modifying the order of magnitude estimates presented here.

Our starting point is the integral in Eq.~\eqref{eq:OmegaGW-gen}. With the same reasoning of that section this integral can be split into two parts, accounting for the amount of time spent by a given mode in the blue or red part of the spectrum. In this case,  however, we have to consider that the solution changes at the transition between the \textit{rolling} and the \textit{minimum phases}, i.e. at $\tau=\tau_{\rm tr}$. This effect can be parametrized  by a power-law peak emission
\begin{equation}
  P_{\rm peak}(\tau)\sim \tau^{-\gamma(\tau)}   \,,
\end{equation}
with the time-dependent function $\gamma(\tau)$ interpolating between the constant peak exponents $\gamma_{\rm r}$ and $\gamma_{\rm m}$ associated to the \textit{rolling} and \textit{minimum phases}, cf. Table \ref{tab:table2}. Assuming, as explained above, a rapid transition between this two phases,
\begin{equation}
 \gamma(\tau)= \gamma_{\rm r}\mathcal{S}(\tau_{\rm tr}-\tau)+\gamma_{\rm m}\mathcal{S}(\tau -\tau_{\rm tr})\,,
\end{equation}
with $\mathcal{S}(\tau)$ the Heaviside step function, the integral in  Eq.~\eqref{eq:OmegaGW-gen} can be written as
\begin{equation}
 \Omega_{\rm GW}(\tau)\sim \frac{P_{\rm peak}(\tau_{\rm kin})}{\Theta}\left(\frac{a(\tau)}{a_{\rm rad}}\right)^2\left[\kappa^\alpha\hspace{-1mm}\int_1^{\kappa^{-1}}ds'\left((s')^{\alpha -2-\gamma(s')}\right) +\kappa^{-\bar\alpha}\hspace{-1mm}\int_{x^{-1}}^{s(\tau)}ds'\left((s')^{-\bar\alpha -2-\gamma(s')}\right) \right]\,,
\end{equation}
where we have specialized to kinetic domination and defined dimensionless conformal times and momenta as in \eqref{eq:s_and_x}.
This integral can be analytically performed to obtain
\begin{equation}
 \Omega_{\rm GW}(\tau)\sim P_{\rm peak}(\tau_{\rm kin})\Theta^{-1}\left(\frac{a(\tau)}{a_{\rm rad}}\right)^2 F(\kappa,s(\tau),s_{\rm tr})\,,
\end{equation}
with the form factor $F$
\begin{equation}
 F(\kappa,s(\tau),s_{\rm tr}) =F_+(\kappa,s_{\rm tr})+F_-(\kappa,s(\tau),s_{\rm tr})
\end{equation}
made of two pieces
\begin{eqnarray}\label{eq:F_plus}
\hspace{-8mm} F_+(\kappa,s_{\rm tr}) &=& \kappa^\alpha
\begin{cases}
\frac{1-\kappa^{\beta_{\rm r}-\alpha}}{\beta_{\rm r}-\alpha} &\quad \text{for} \quad s_{\rm tr}\,\kappa>1\\
\frac{1-s_{\rm tr}^{-\beta_{\rm r}+\alpha}}{\beta_{\rm r}-\alpha}&\quad \text{for} \quad s_{\rm tr}\,\kappa=1\\
\frac{ s_{\rm tr}^{1-\beta_{\rm r}-\beta_{\rm m}}}{(\beta_{\rm m}-\alpha)(\alpha-\beta_{\rm r})}\left[ s_{\rm tr}^{\alpha-1}(s_{\rm tr}^{\beta_{\rm m}}(\beta_{\rm m}-\alpha)- s_{\rm tr}^{\beta_{\rm r}}(\beta_{\rm r}-\alpha))\right.\\
\left.\qquad+s_{\rm tr}^{\beta_{\rm r}+\beta_{\rm m}-1}(\kappa^{\beta_{\rm m}-\alpha}(\beta_{\rm r}-\alpha)-(\beta_{\rm m}-\alpha))\right] &\quad \text{for} \quad s_{\rm tr}\,\kappa<1
\end{cases}
\end{eqnarray}

\begin{eqnarray}\label{eq:F_min}
\hspace{-8mm} F_-(\kappa,s(\tau),s_{\rm tr}) &=& \kappa^{-\bar\alpha}
\begin{cases}
\frac{s_{\rm tr}^{-\bar\alpha-\beta_{\rm m}-\beta_{\rm r}+1}}{(\bar\alpha+\beta_{\rm m})(\bar\alpha+\beta_{\rm r})}\left[s_{\rm tr}^{\beta_{\rm m}-1}(\bar\alpha+\beta_{\rm m})(-1+s_{\rm tr}^{\bar\alpha+\beta_{\rm r}}\kappa^{\bar\alpha+\beta_{\rm r}})\right.\\
\left.\qquad+(\bar\alpha+\beta_{\rm r})s_{\rm tr}^{\beta_{\rm r}-1}(1-s_{\rm tr}^{\bar\alpha+\beta_{\rm m}}s(\tau)^{-\bar\alpha-\beta_{\rm m}})\right] &\quad \text{for} \quad s_{\rm tr}\,\kappa>1\\
\frac{s_{\rm tr}^{-\bar\alpha-\beta_{\rm m}}-s(\tau)^{-\bar\alpha-\beta_{\rm m}}}{\bar\alpha+\beta_{\rm m}}&\quad \text{for} \quad s_{\rm tr}\,\kappa=1\\
\frac{\kappa^{\bar\alpha+\beta_{\rm m}}-s(\tau)^{-\bar\alpha-\beta_{\rm m}}}{\bar\alpha+\beta_{\rm m}}&\quad \text{for} \quad s_{\rm tr}\,\kappa<1

\end{cases}
\end{eqnarray}
We have confronted this involved analytical result with the numerical integration of the spectrum using a smoother transition function   between the \textit{rolling} and \textit{minimum phases}, finding a very good qualitative agreement. The main difference between the two approaches is the time needed for the field reach the minimum, being this one slightly larger in the sudden approximation.

\bibliographystyle{JHEP.bst}

\bibliography{cosmicbibliography_main.bib}

\providecommand{\href}[2]{#2}\begingroup\raggedright\begin{thebibliography}{10}

\bibitem{Abbott:2016blz}
{\scshape Virgo, LIGO Scientific} collaboration, B.~P. Abbott et~al.,
  \emph{{Observation of Gravitational Waves from a Binary Black Hole Merger}},
  \href{https://doi.org/10.1103/PhysRevLett.116.061102}{\emph{Phys. Rev. Lett.}
  {\bfseries 116} (2016) 061102}
  [\href{https://arxiv.org/abs/1602.03837}{{\ttfamily 1602.03837}}].

\bibitem{Abbott:2016nmj}
{\scshape Virgo, LIGO Scientific} collaboration, B.~P. Abbott et~al.,
  \emph{{GW151226: Observation of Gravitational Waves from a 22-Solar-Mass
  Binary Black Hole Coalescence}},
  \href{https://doi.org/10.1103/PhysRevLett.116.241103}{\emph{Phys. Rev. Lett.}
  {\bfseries 116} (2016) 241103}
  [\href{https://arxiv.org/abs/1606.04855}{{\ttfamily 1606.04855}}].

\bibitem{TheLIGOScientific:2017qsa}
{\scshape Virgo, LIGO Scientific} collaboration, B.~P. Abbott et~al.,
  \emph{{GW170817: Observation of Gravitational Waves from a Binary Neutron
  Star Inspiral}},
  \href{https://doi.org/10.1103/PhysRevLett.119.161101}{\emph{Phys. Rev. Lett.}
  {\bfseries 119} (2017) 161101}
  [\href{https://arxiv.org/abs/1710.05832}{{\ttfamily 1710.05832}}].

\bibitem{Monitor:2017mdv}
{\scshape Virgo, Fermi-GBM, INTEGRAL, LIGO Scientific} collaboration, B.~P.
  Abbott et~al., \emph{{Gravitational Waves and Gamma-rays from a Binary
  Neutron Star Merger: GW170817 and GRB 170817A}},
  \href{https://doi.org/10.3847/2041-8213/aa920c}{\emph{Astrophys. J.}
  {\bfseries 848} (2017) L13}
  [\href{https://arxiv.org/abs/1710.05834}{{\ttfamily 1710.05834}}].

\bibitem{Starobinsky:1979ty}
A.~A. Starobinsky, \emph{{Spectrum of relict gravitational radiation and the
  early state of the universe}}, {\emph{JETP Lett.} {\bfseries 30} (1979) 682}.

\bibitem{Rubakov:1982df}
V.~A. Rubakov, M.~V. Sazhin and A.~V. Veryaskin, \emph{{Graviton Creation in
  the Inflationary Universe and the Grand Unification Scale}},
  \href{https://doi.org/10.1016/0370-2693(82)90641-4}{\emph{Phys. Lett.}
  {\bfseries 115B} (1982) 189}.

\bibitem{GarciaBellido:2007af}
J.~Garcia-Bellido, D.~G. Figueroa and A.~Sastre, \emph{{A Gravitational Wave
  Background from Reheating after Hybrid Inflation}},
  \href{https://doi.org/10.1103/PhysRevD.77.043517}{\emph{Phys. Rev.}
  {\bfseries D77} (2008) 043517}
  [\href{https://arxiv.org/abs/0707.0839}{{\ttfamily 0707.0839}}].

\bibitem{Grojean:2006bp}
C.~Grojean and G.~Servant, \emph{{Gravitational Waves from Phase Transitions at
  the Electroweak Scale and Beyond}},
  \href{https://doi.org/10.1103/PhysRevD.75.043507}{\emph{Phys. Rev.}
  {\bfseries D75} (2007) 043507}
  [\href{https://arxiv.org/abs/hep-ph/0607107}{{\ttfamily hep-ph/0607107}}].

\bibitem{Weir:2017wfa}
D.~J. Weir, \emph{{Gravitational waves from a first order electroweak phase
  transition: a brief review}},
  \href{https://doi.org/10.1098/rsta.2017.0126}{\emph{Phil. Trans. Roy. Soc.
  Lond.} {\bfseries A376} (2018) 20170126}
  [\href{https://arxiv.org/abs/1705.01783}{{\ttfamily 1705.01783}}].

\bibitem{Vilenkin:2000jqa}
A.~Vilenkin and E.~P.~S. Shellard, \emph{{Cosmic Strings and Other Topological
  Defects}}. Cambridge University Press, 2000.

\bibitem{Copeland:2009ga}
E.~J. Copeland and T.~W.~B. Kibble, \emph{{Cosmic Strings and Superstrings}},
  \href{https://doi.org/10.1098/rspa.2009.0591}{\emph{Proc. Roy. Soc. Lond.}
  {\bfseries A466} (2010) 623}
  [\href{https://arxiv.org/abs/0911.1345}{{\ttfamily 0911.1345}}].

\bibitem{Dufaux:2010cf}
J.-F. Dufaux, D.~G. Figueroa and J.~Garcia-Bellido, \emph{{Gravitational Waves
  from Abelian Gauge Fields and Cosmic Strings at Preheating}},
  \href{https://doi.org/10.1103/PhysRevD.82.083518}{\emph{Phys. Rev.}
  {\bfseries D82} (2010) 083518}
  [\href{https://arxiv.org/abs/1006.0217}{{\ttfamily 1006.0217}}].

\bibitem{Figueroa:2012kw}
D.~G. Figueroa, M.~Hindmarsh and J.~Urrestilla, \emph{{Exact Scale-Invariant
  Background of Gravitational Waves from Cosmic Defects}},
  \href{https://doi.org/10.1103/PhysRevLett.110.101302}{\emph{Phys. Rev. Lett.}
  {\bfseries 110} (2013) 101302}
  [\href{https://arxiv.org/abs/1212.5458}{{\ttfamily 1212.5458}}].

\bibitem{Bettoni:2018utf}
D.~Bettoni and J.~Rubio, \emph{{Quintessential Affleck-Dine baryogenesis with
  non-minimal couplings}},
  \href{https://doi.org/10.1016/j.physletb.2018.07.046}{\emph{Phys. Lett.}
  {\bfseries B784} (2018) 122}
  [\href{https://arxiv.org/abs/1805.02669}{{\ttfamily 1805.02669}}].

\bibitem{GarciaBellido:2011de}
J.~Garcia-Bellido, J.~Rubio, M.~Shaposhnikov and D.~Zenhausern,
  \emph{{Higgs-Dilaton Cosmology: From the Early to the Late Universe}},
  \href{https://doi.org/10.1103/PhysRevD.84.123504}{\emph{Phys. Rev.}
  {\bfseries D84} (2011) 123504}
  [\href{https://arxiv.org/abs/1107.2163}{{\ttfamily 1107.2163}}].

\bibitem{Karananas:2016kyt}
G.~K. Karananas and J.~Rubio, \emph{{On the geometrical interpretation of
  scale-invariant models of inflation}},
  \href{https://doi.org/10.1016/j.physletb.2016.08.037}{\emph{Phys. Lett.}
  {\bfseries B761} (2016) 223}
  [\href{https://arxiv.org/abs/1606.08848}{{\ttfamily 1606.08848}}].

\bibitem{Casas:2017wjh}
S.~Casas, M.~Pauly and J.~Rubio, \emph{{Higgs-dilaton cosmology: An
  inflation–dark-energy connection and forecasts for future galaxy surveys}},
  \href{https://doi.org/10.1103/PhysRevD.97.043520}{\emph{Phys. Rev.}
  {\bfseries D97} (2018) 043520}
  [\href{https://arxiv.org/abs/1712.04956}{{\ttfamily 1712.04956}}].

\bibitem{Wetterich:1987fm}
C.~Wetterich, \emph{{Cosmology and the Fate of Dilatation Symmetry}},
  \href{https://doi.org/10.1016/0550-3213(88)90193-9}{\emph{Nucl. Phys.}
  {\bfseries B302} (1988) 668}
  [\href{https://arxiv.org/abs/1711.03844}{{\ttfamily 1711.03844}}].

\bibitem{Wetterich:1994bg}
C.~Wetterich, \emph{{The Cosmon model for an asymptotically vanishing time
  dependent cosmological 'constant'}}, {\emph{Astron. Astrophys.} {\bfseries
  301} (1995) 321} [\href{https://arxiv.org/abs/hep-th/9408025}{{\ttfamily
  hep-th/9408025}}].

\bibitem{Wetterich:2014gaa}
C.~Wetterich, \emph{{Inflation, quintessence, and the origin of mass}},
  \href{https://doi.org/10.1016/j.nuclphysb.2015.05.019}{\emph{Nucl. Phys.}
  {\bfseries B897} (2015) 111}
  [\href{https://arxiv.org/abs/1408.0156}{{\ttfamily 1408.0156}}].

\bibitem{Rubio:2017gty}
J.~Rubio and C.~Wetterich, \emph{{Emergent scale symmetry: Connecting inflation
  and dark energy}},
  \href{https://doi.org/10.1103/PhysRevD.96.063509}{\emph{Phys. Rev.}
  {\bfseries D96} (2017) 063509}
  [\href{https://arxiv.org/abs/1705.00552}{{\ttfamily 1705.00552}}].

\bibitem{Peebles:1998qn}
P.~J.~E. Peebles and A.~Vilenkin, \emph{{Quintessential inflation}},
  \href{https://doi.org/10.1103/PhysRevD.59.063505}{\emph{Phys. Rev.}
  {\bfseries D59} (1999) 063505}
  [\href{https://arxiv.org/abs/astro-ph/9810509}{{\ttfamily
  astro-ph/9810509}}].

\bibitem{Spokoiny:1993kt}
B.~Spokoiny, \emph{{Deflationary universe scenario}},
  \href{https://doi.org/10.1016/0370-2693(93)90155-B}{\emph{Phys. Lett.}
  {\bfseries B315} (1993) 40}
  [\href{https://arxiv.org/abs/gr-qc/9306008}{{\ttfamily gr-qc/9306008}}].

\bibitem{Brax:2005uf}
P.~Brax and J.~Martin, \emph{{Coupling quintessence to inflation in
  supergravity}}, \href{https://doi.org/10.1103/PhysRevD.71.063530}{\emph{Phys.
  Rev.} {\bfseries D71} (2005) 063530}
  [\href{https://arxiv.org/abs/astro-ph/0502069}{{\ttfamily
  astro-ph/0502069}}].

\bibitem{Hossain:2014xha}
M.~W. Hossain, R.~Myrzakulov, M.~Sami and E.~N. Saridakis, \emph{{Variable
  gravity: A suitable framework for quintessential inflation}},
  \href{https://doi.org/10.1103/PhysRevD.90.023512}{\emph{Phys. Rev.}
  {\bfseries D90} (2014) 023512}
  [\href{https://arxiv.org/abs/1402.6661}{{\ttfamily 1402.6661}}].

\bibitem{Agarwal:2017wxo}
A.~Agarwal, R.~Myrzakulov, M.~Sami and N.~K. Singh, \emph{{Quintessential
  inflation in a thawing realization}},
  \href{https://doi.org/10.1016/j.physletb.2017.04.066}{\emph{Phys. Lett.}
  {\bfseries B770} (2017) 200}
  [\href{https://arxiv.org/abs/1708.00156}{{\ttfamily 1708.00156}}].

\bibitem{Geng:2017mic}
C.-Q. Geng, C.-C. Lee, M.~Sami, E.~N. Saridakis and A.~A. Starobinsky,
  \emph{{Observational constraints on successful model of quintessential
  Inflation}}, \href{https://doi.org/10.1088/1475-7516/2017/06/011}{\emph{JCAP}
  {\bfseries 1706} (2017) 011}
  [\href{https://arxiv.org/abs/1705.01329}{{\ttfamily 1705.01329}}].

\bibitem{Dimopoulos:2017zvq}
K.~Dimopoulos and C.~Owen, \emph{{Quintessential Inflation with
  $\alpha$-attractors}},
  \href{https://doi.org/10.1088/1475-7516/2017/06/027}{\emph{JCAP} {\bfseries
  1706} (2017) 027} [\href{https://arxiv.org/abs/1703.00305}{{\ttfamily
  1703.00305}}].

\bibitem{Peccei:1987mm}
R.~D. Peccei, J.~Sola and C.~Wetterich, \emph{{Adjusting the Cosmological
  Constant Dynamically: Cosmons and a New Force Weaker Than Gravity}},
  \href{https://doi.org/10.1016/0370-2693(87)91191-9}{\emph{Phys. Lett.}
  {\bfseries B195} (1987) 183}.

\bibitem{Wetterich:2013jsa}
C.~Wetterich, \emph{{Variable gravity Universe}},
  \href{https://doi.org/10.1103/PhysRevD.89.024005}{\emph{Phys. Rev.}
  {\bfseries D89} (2014) 024005}
  [\href{https://arxiv.org/abs/1308.1019}{{\ttfamily 1308.1019}}].

\bibitem{Wang:2018kly}
S.-J. Wang, \emph{{Quintessential Starobinsky inflation and swampland
  criteria}},  \href{https://arxiv.org/abs/1810.06445}{{\ttfamily 1810.06445}}.

\bibitem{Ford:1986sy}
L.~H. Ford, \emph{{Gravitational Particle Creation and Inflation}},
  \href{https://doi.org/10.1103/PhysRevD.35.2955}{\emph{Phys. Rev.} {\bfseries
  D35} (1987) 2955}.

\bibitem{Damour:1995pd}
T.~Damour and A.~Vilenkin, \emph{{String theory and inflation}},
  \href{https://doi.org/10.1103/PhysRevD.53.2981}{\emph{Phys. Rev.} {\bfseries
  D53} (1996) 2981} [\href{https://arxiv.org/abs/hep-th/9503149}{{\ttfamily
  hep-th/9503149}}].

\bibitem{Felder:1999pv}
G.~N. Felder, L.~Kofman and A.~D. Linde, \emph{{Inflation and preheating in NO
  models}}, \href{https://doi.org/10.1103/PhysRevD.60.103505}{\emph{Phys. Rev.}
  {\bfseries D60} (1999) 103505}
  [\href{https://arxiv.org/abs/hep-ph/9903350}{{\ttfamily hep-ph/9903350}}].

\bibitem{Feng:2002nb}
B.~Feng and M.-z. Li, \emph{{Curvaton reheating in nonoscillatory inflationary
  models}}, \href{https://doi.org/10.1016/S0370-2693(03)00589-6}{\emph{Phys.
  Lett.} {\bfseries B564} (2003) 169}
  [\href{https://arxiv.org/abs/hep-ph/0212213}{{\ttfamily hep-ph/0212213}}].

\bibitem{BuenoSanchez:2007jxm}
J.~C. Bueno~Sanchez and K.~Dimopoulos, \emph{{Curvaton reheating allows TeV
  Hubble scale in NO inflation}},
  \href{https://doi.org/10.1088/1475-7516/2007/11/007}{\emph{JCAP} {\bfseries
  0711} (2007) 007} [\href{https://arxiv.org/abs/0707.3967}{{\ttfamily
  0707.3967}}].

\bibitem{Dimopoulos:2018wfg}
K.~Dimopoulos and T.~Markkanen, \emph{{Non-minimal gravitational reheating
  during kination}},
  \href{https://doi.org/10.1088/1475-7516/2018/06/021}{\emph{JCAP} {\bfseries
  1806} (2018) 021} [\href{https://arxiv.org/abs/1803.07399}{{\ttfamily
  1803.07399}}].

\bibitem{Nakama:2018gll}
T.~Nakama and J.~Yokoyama, \emph{{Reheating through the Higgs amplified by
  spinodal instabilities and gravitational creation of gravitons}},
  \href{https://arxiv.org/abs/1803.07111}{{\ttfamily 1803.07111}}.

\bibitem{Sahni:1990tx}
V.~Sahni, \emph{{The Energy Density of Relic Gravity Waves From Inflation}},
  \href{https://doi.org/10.1103/PhysRevD.42.453}{\emph{Phys. Rev.} {\bfseries
  D42} (1990) 453}.

\bibitem{Maggiore:1999vm}
M.~Maggiore, \emph{{Gravitational wave experiments and early universe
  cosmology}}, \href{https://doi.org/10.1016/S0370-1573(99)00102-7}{\emph{Phys.
  Rept.} {\bfseries 331} (2000) 283}
  [\href{https://arxiv.org/abs/gr-qc/9909001}{{\ttfamily gr-qc/9909001}}].

\bibitem{Caprini:2018mtu}
C.~Caprini and D.~G. Figueroa, \emph{{Cosmological Backgrounds of Gravitational
  Waves}}, \href{https://doi.org/10.1088/1361-6382/aac608}{\emph{Class. Quant.
  Grav.} {\bfseries 35} (2018) 163001}
  [\href{https://arxiv.org/abs/1801.04268}{{\ttfamily 1801.04268}}].

\bibitem{Wetterich:2007kr}
C.~Wetterich, \emph{{Growing neutrinos and cosmological selection}},
  \href{https://doi.org/10.1016/j.physletb.2007.08.060}{\emph{Phys. Lett.}
  {\bfseries B655} (2007) 201}
  [\href{https://arxiv.org/abs/0706.4427}{{\ttfamily 0706.4427}}].

\bibitem{Amendola:2018ltt}
L.~Amendola, D.~Bettoni, G.~Domènech and A.~R. Gomes, \emph{{Doppelg\"anger
  dark energy: modified gravity with non-universal couplings after GW170817}},
  \href{https://doi.org/10.1088/1475-7516/2018/06/029}{\emph{JCAP} {\bfseries
  1806} (2018) 029} [\href{https://arxiv.org/abs/1803.06368}{{\ttfamily
  1803.06368}}].

\bibitem{Amendola:2007yx}
L.~Amendola, M.~Baldi and C.~Wetterich, \emph{{Quintessence cosmologies with a
  growing matter component}},
  \href{https://doi.org/10.1103/PhysRevD.78.023015}{\emph{Phys. Rev.}
  {\bfseries D78} (2008) 023015}
  [\href{https://arxiv.org/abs/0706.3064}{{\ttfamily 0706.3064}}].

\bibitem{Birrell:1982ix}
N.~D. Birrell and P.~C.~W. Davies, \emph{{Quantum Fields in Curved Space}},
  Cambridge Monographs on Mathematical Physics. Cambridge Univ. Press,
  Cambridge, UK, 1984,
  \href{https://doi.org/10.1017/CBO9780511622632}{10.1017/CBO9780511622632}.

\bibitem{Fairbairn:2018bsw}
M.~Fairbairn, K.~Kainulainen, T.~Markkanen and S.~Nurmi, \emph{{Despicable Dark
  Relics: generated by gravity with unconstrained masses}},
  \href{https://arxiv.org/abs/1808.08236}{{\ttfamily 1808.08236}}.

\bibitem{Alonso-Alvarez:2018tus}
G.~Alonso-Álvarez and J.~Jaeckel, \emph{{Lightish but clumpy: scalar dark
  matter from inflationary fluctuations}},
  \href{https://doi.org/10.1088/1475-7516/2018/10/022}{\emph{JCAP} {\bfseries
  1810} (2018) 022} [\href{https://arxiv.org/abs/1807.09785}{{\ttfamily
  1807.09785}}].

\bibitem{Zeldovich:1974uw}
{\relax Ya}.~B. Zeldovich, I.~{\relax Yu}. Kobzarev and L.~B. Okun,
  \emph{{Cosmological Consequences of the Spontaneous Breakdown of Discrete
  Symmetry}}, {\emph{Zh. Eksp. Teor. Fiz.} {\bfseries 67} (1974) 3}.

\bibitem{Kibble:1976sj}
T.~W.~B. Kibble, \emph{{Topology of Cosmic Domains and Strings}},
  \href{https://doi.org/10.1088/0305-4470/9/8/029}{\emph{J. Phys.} {\bfseries
  A9} (1976) 1387}.

\bibitem{Hindmarsh:1994re}
M.~B. Hindmarsh and T.~W.~B. Kibble, \emph{{Cosmic strings}},
  \href{https://doi.org/10.1088/0034-4885/58/5/001}{\emph{Rept. Prog. Phys.}
  {\bfseries 58} (1995) 477}
  [\href{https://arxiv.org/abs/hep-ph/9411342}{{\ttfamily hep-ph/9411342}}].

\bibitem{Moore:2016itg}
G.~D. Moore, \emph{{Intercommutation of U(1) global cosmic strings}},
  \href{https://arxiv.org/abs/1604.02356}{{\ttfamily 1604.02356}}.

\bibitem{Gorghetto:2018myk}
M.~Gorghetto, E.~Hardy and G.~Villadoro, \emph{{Axions from Strings: the
  Attractive Solution}},
  \href{https://doi.org/10.1007/JHEP07(2018)151}{\emph{JHEP} {\bfseries 07}
  (2018) 151} [\href{https://arxiv.org/abs/1806.04677}{{\ttfamily
  1806.04677}}].

\bibitem{Vaquero:2018tib}
A.~Vaquero, J.~Redondo and J.~Stadler, \emph{{Early seeds of of axion
  miniclusters}},  \href{https://arxiv.org/abs/1809.09241}{{\ttfamily
  1809.09241}}.

\bibitem{Damour:2001bk}
T.~Damour and A.~Vilenkin, \emph{{Gravitational wave bursts from cusps and
  kinks on cosmic strings}},
  \href{https://doi.org/10.1103/PhysRevD.64.064008}{\emph{Phys. Rev.}
  {\bfseries D64} (2001) 064008}
  [\href{https://arxiv.org/abs/gr-qc/0104026}{{\ttfamily gr-qc/0104026}}].

\bibitem{Siemens:2006yp}
X.~Siemens, V.~Mandic and J.~Creighton, \emph{{Gravitational wave stochastic
  background from cosmic (super)strings}},
  \href{https://doi.org/10.1103/PhysRevLett.98.111101}{\emph{Phys. Rev. Lett.}
  {\bfseries 98} (2007) 111101}
  [\href{https://arxiv.org/abs/astro-ph/0610920}{{\ttfamily
  astro-ph/0610920}}].

\bibitem{Kawasaki:2011dp}
M.~Kawasaki, K.~Miyamoto and K.~Nakayama, \emph{{Cosmological Effects of
  Decaying Cosmic String Loops with TeV Scale Width}},
  \href{https://arxiv.org/abs/1105.4383}{{\ttfamily 1105.4383}}.

\bibitem{Bennett:1989ak}
D.~P. Bennett and F.~R. Bouchet, \emph{{Cosmic string evolution}},
  \href{https://doi.org/10.1103/PhysRevLett.63.2776}{\emph{Phys. Rev. Lett.}
  {\bfseries 63} (1989) 2776}.

\bibitem{Allen:1990tv}
B.~Allen and E.~P.~S. Shellard, \emph{{Cosmic string evolution: a numerical
  simulation}}, \href{https://doi.org/10.1103/PhysRevLett.64.119}{\emph{Phys.
  Rev. Lett.} {\bfseries 64} (1990) 119}.

\bibitem{BlancoPillado:2011dq}
J.~J. Blanco-Pillado, K.~D. Olum and B.~Shlaer, \emph{{Large parallel cosmic
  string simulations: New results on loop production}},
  \href{https://doi.org/10.1103/PhysRevD.83.083514}{\emph{Phys. Rev.}
  {\bfseries D83} (2011) 083514}
  [\href{https://arxiv.org/abs/1101.5173}{{\ttfamily 1101.5173}}].

\bibitem{Ringeval:2005kr}
C.~Ringeval, M.~Sakellariadou and F.~Bouchet, \emph{{Cosmological evolution of
  cosmic string loops}},
  \href{https://doi.org/10.1088/1475-7516/2007/02/023}{\emph{JCAP} {\bfseries
  0702} (2007) 023} [\href{https://arxiv.org/abs/astro-ph/0511646}{{\ttfamily
  astro-ph/0511646}}].

\bibitem{Krauss:1991qu}
L.~M. Krauss, \emph{{Gravitational waves from global phase transitions}},
  \href{https://doi.org/10.1016/0370-2693(92)90425-4}{\emph{Phys. Lett.}
  {\bfseries B284} (1992) 229}.

\bibitem{Kamada:2015iga}
A.~Kamada and M.~Yamada, \emph{{Gravitational wave signals from short-lived
  topological defects in the MSSM}},
  \href{https://doi.org/10.1088/1475-7516/2015/10/021}{\emph{JCAP} {\bfseries
  1510} (2015) 021} [\href{https://arxiv.org/abs/1505.01167}{{\ttfamily
  1505.01167}}].

\bibitem{Dufaux:2007pt}
J.~F. Dufaux, A.~Bergman, G.~N. Felder, L.~Kofman and J.-P. Uzan, \emph{{Theory
  and Numerics of Gravitational Waves from Preheating after Inflation}},
  \href{https://doi.org/10.1103/PhysRevD.76.123517}{\emph{Phys. Rev.}
  {\bfseries D76} (2007) 123517}
  [\href{https://arxiv.org/abs/0707.0875}{{\ttfamily 0707.0875}}].

\bibitem{Kamada:2014qja}
A.~Kamada and M.~Yamada, \emph{{Gravitational waves as a probe of the SUSY
  scale}}, \href{https://doi.org/10.1103/PhysRevD.91.063529}{\emph{Phys. Rev.}
  {\bfseries D91} (2015) 063529}
  [\href{https://arxiv.org/abs/1407.2882}{{\ttfamily 1407.2882}}].

\bibitem{Thrane:2013oya}
E.~Thrane and J.~D. Romano, \emph{{Sensitivity curves for searches for
  gravitational-wave backgrounds}},
  \href{https://doi.org/10.1103/PhysRevD.88.124032}{\emph{Phys. Rev.}
  {\bfseries D88} (2013) 124032}
  [\href{https://arxiv.org/abs/1310.5300}{{\ttfamily 1310.5300}}].

\bibitem{Breitbach:2018ddu}
M.~Breitbach, J.~Kopp, E.~Madge, T.~Opferkuch and P.~Schwaller, \emph{{Dark,
  Cold, and Noisy: Constraining Secluded Hidden Sectors with Gravitational
  Waves}},  \href{https://arxiv.org/abs/1811.11175}{{\ttfamily 1811.11175}}.

\bibitem{adv_ligo}
\url{https://dcc.ligo.org/}.

\bibitem{virgo}
\url{http://www.virgo-gw.eu/}.

\bibitem{et}
\url{http://www.et-gw.eu}.

\bibitem{lisa}
\url{http://lisa.nasa.gov/}.

\bibitem{decigo}
\url{http://tamago.mtk.nao.ac.jp/decigo/index_E.html}.

\bibitem{Yokoyama:1988zza}
J.~Yokoyama, \emph{{Natural Way Out of the Conflict Between Cosmic Strings and
  Inflation}}, \href{https://doi.org/10.1016/0370-2693(88)91316-0}{\emph{Phys.
  Lett.} {\bfseries B212} (1988) 273}.

\bibitem{Hagmann:1990mj}
C.~Hagmann and P.~Sikivie, \emph{{Computer simulations of the motion and decay
  of global strings}},
  \href{https://doi.org/10.1016/0550-3213(91)90243-Q}{\emph{Nucl. Phys.}
  {\bfseries B363} (1991) 247}.

\bibitem{Yamaguchi:1998gx}
M.~Yamaguchi, M.~Kawasaki and J.~Yokoyama, \emph{{Evolution of axionic strings
  and spectrum of axions radiated from them}},
  \href{https://doi.org/10.1103/PhysRevLett.82.4578}{\emph{Phys. Rev. Lett.}
  {\bfseries 82} (1999) 4578}
  [\href{https://arxiv.org/abs/hep-ph/9811311}{{\ttfamily hep-ph/9811311}}].

\bibitem{Yamaguchi:1999yp}
M.~Yamaguchi, \emph{{Scaling property of the global string in the radiation
  dominated universe}},
  \href{https://doi.org/10.1103/PhysRevD.60.103511}{\emph{Phys. Rev.}
  {\bfseries D60} (1999) 103511}
  [\href{https://arxiv.org/abs/hep-ph/9907506}{{\ttfamily hep-ph/9907506}}].

\bibitem{Yamaguchi:1999dy}
M.~Yamaguchi, J.~Yokoyama and M.~Kawasaki, \emph{{Evolution of a global string
  network in a matter dominated universe}},
  \href{https://doi.org/10.1103/PhysRevD.61.061301}{\emph{Phys. Rev.}
  {\bfseries D61} (2000) 061301}
  [\href{https://arxiv.org/abs/hep-ph/9910352}{{\ttfamily hep-ph/9910352}}].

\bibitem{Cui:2017ufi}
Y.~Cui, M.~Lewicki, D.~E. Morrissey and J.~D. Wells, \emph{{Cosmic Archaeology
  with Gravitational Waves from Cosmic Strings}},
  \href{https://doi.org/10.1103/PhysRevD.97.123505}{\emph{Phys. Rev.}
  {\bfseries D97} (2018) 123505}
  [\href{https://arxiv.org/abs/1711.03104}{{\ttfamily 1711.03104}}].

\bibitem{Cui:2018rwi}
Y.~Cui, M.~Lewicki, D.~E. Morrissey and J.~D. Wells, \emph{{Probing the pre-BBN
  universe with gravitational waves from cosmic strings}},
  \href{https://arxiv.org/abs/1808.08968}{{\ttfamily 1808.08968}}.

\bibitem{Artymowski:2017pua}
M.~Artymowski, O.~Czerwinska, Z.~Lalak and M.~Lewicki, \emph{{Gravitational
  wave signals and cosmological consequences of gravitational reheating}},
  \href{https://doi.org/10.1088/1475-7516/2018/04/046}{\emph{JCAP} {\bfseries
  1804} (2018) 046} [\href{https://arxiv.org/abs/1711.08473}{{\ttfamily
  1711.08473}}].

\bibitem{Dimopoulos:2017tud}
K.~Dimopoulos, L.~Donaldson~Wood and C.~Owen, \emph{{Instant preheating in
  quintessential inflation with $\alpha$-attractors}},
  \href{https://doi.org/10.1103/PhysRevD.97.063525}{\emph{Phys. Rev.}
  {\bfseries D97} (2018) 063525}
  [\href{https://arxiv.org/abs/1712.01760}{{\ttfamily 1712.01760}}].

\bibitem{Akrami:2017cir}
Y.~Akrami, R.~Kallosh, A.~Linde and V.~Vardanyan, \emph{{Dark energy,
  $\alpha$-attractors, and large-scale structure surveys}},
  \href{https://doi.org/10.1088/1475-7516/2018/06/041}{\emph{JCAP} {\bfseries
  1806} (2018) 041} [\href{https://arxiv.org/abs/1712.09693}{{\ttfamily
  1712.09693}}].

\bibitem{Garcia-Garcia:2018hlc}
C.~Garcia-Garcia, E.~V. Linder, P.~Ruiz-Lapuente and M.~Zumalacarregui,
  \emph{{Dark energy from $\alpha$-attractors: phenomenology and observational
  constraints}},
  \href{https://doi.org/10.1088/1475-7516/2018/08/022}{\emph{JCAP} {\bfseries
  1808} (2018) 022} [\href{https://arxiv.org/abs/1803.00661}{{\ttfamily
  1803.00661}}].

\bibitem{Amendola:2017xhl}
L.~Amendola, J.~Rubio and C.~Wetterich, \emph{{Primordial black holes from
  fifth forces}}, \href{https://doi.org/10.1103/PhysRevD.97.081302}{\emph{Phys.
  Rev.} {\bfseries D97} (2018) 081302}
  [\href{https://arxiv.org/abs/1711.09915}{{\ttfamily 1711.09915}}].

\end{thebibliography}\endgroup

\end{document}